\documentclass[usenatbib]{mn2e}
\usepackage{graphicx,fixltx2e}

\def\chisqr{\hbox{$\chi^2_{\rm r}$}}
\def\msun{\hbox{${\rm M}_{\odot}$}}
\def\mspy{\hbox{${\rm M}_{\odot}$\,yr$^{-1}$}}
\def\rsun{\hbox{${\rm R}_{\odot}$}}
\def\rcor{\hbox{$r_{\rm C}$}}
\def\rmag{\hbox{$r_{\rm A}$}}

\def\rstar{\hbox{$R_{\star}$}}
\def\teff{\hbox{$T_{\rm eff}$}}
\def\logg{\hbox{$\log g$}}
\def\sn{\hbox{S/N}}
\def\vrad{\hbox{$v_{\rm rad}$}}

\def\kms{\hbox{km\,s$^{-1}$}}
\def\vsini{\hbox{$v \sin i$}}

\def\ptt{\hbox{$10^{-4} I_{\rm c}$}}

\def\degr{\hbox{$^\circ$}}

\def\Mdot{\hbox{$\dot{M}$}}

\def\Ip{\hbox{$I_{\rm p}$}}
\def\Im{\hbox{$I_{\rm m}$}}
\def\Ia{\hbox{$I_{\rm a}$}}
\def\Vm{\hbox{$V_{\rm m}$}}
\def\Va{\hbox{$V_{\rm a}$}}
\def\Iq{\hbox{$I_{\rm q}$}}
\def\Vp{\hbox{$V_{\rm p}$}}
\def\Ic{\hbox{$I_{\rm c}$}}
\def\Ie{\hbox{$I_{\rm e}$}}
\def\Ve{\hbox{$V_{\rm e}$}}

\newcommand{\caii}{\hbox{Ca$\;${\sc ii}}}
\newcommand{\fei}{\hbox{Fe$\;${\sc i}}}
\newcommand{\feii}{\hbox{Fe$\;${\sc ii}}}
\newcommand{\hei}{\hbox{He$\;${\sc i}}}
\newcommand{\hal}{\hbox{H${\alpha}$}}
\newcommand{\hbe}{\hbox{H${\beta}$}}

\begin{document}

\title[Magnetospheric accretion on the cTTS BP Tau]{Magnetospheric accretion 
on the T~Tauri star BP~Tauri\thanks{Based on observations 
obtained at the Canada-France-Hawaii Telescope (CFHT) and at the T\'elescope Bernard Lyot (TBL).  
CFHT is operated by the National Research Council of Canada, the Institut National des Sciences 
de l'Univers of the Centre National de la Recherche Scientifique of France (INSU/CNRS), and the 
University of Hawaii, while TBL is operated by CNRS/INSU.  } }

\makeatletter

\def\newauthor{%
  \end{author@tabular}\par
  \begin{author@tabular}[t]{@{}l@{}}}
\makeatother
 
\author[J.-F.~Donati et al.]
{\vspace{1.7mm}
J.-F.~Donati$^1$\thanks{E-mail: donati@ast.obs-mip.fr (J-FD); 
mmj@st-andrews.ac.uk (MMJ); 
sg64@st-andrews.ac.uk (SGG); 
petit@ast.obs-mip.fr (PP); 
fpaletou@ast.obs-mip.fr (FP); 
jerome.bouvier@obs.ujf-grenoble.fr (JB); 
catherine.dougados@obs.ujf-grenoble.fr (CD); 
francois.menard@obs.ujf-grenoble.fr (FM); 
acc4@st-andrews.ac.uk (ACC); 
th@astro.ex.ac.uk (TJH); 
ghussain@eso.org (GAJH); 
y.unruh@imperial.ac.uk (YU); 
jmorin@ast.obs-mip.fr (JM); 
scm@aao.gov.au (SCM); 
manset@cfht.hawaii.edu (NM); 
auriere@ast.obs-mip.fr (MA); 
claude.catala@obspm.fr (CC);  
evelyne.alecian@obspm.fr (EA) 
}, 
M.M.~Jardine$^2$, S.G.~Gregory$^2$, P.~Petit$^1$, F.~Paletou$^1$, J.~Bouvier$^3$, \\ 
\vspace{1.7mm}
{\hspace{-1.5mm}\LARGE\rm 
C.~Dougados$^3$, F.~M\'enard$^3$, A.C.~Cameron$^2$, T.J.~Harries$^4$, G.A.J.~Hussain$^5$, } \\
\vspace{1.7mm}
{\hspace{-1.5mm}\LARGE\rm
Y.~Unruh$^6$, J.~Morin$^1$, S.C.~Marsden$^7$, N.~Manset$^8$, M.~Auri\`ere$^1$, C.~Catala$^9$, } \\
\vspace{1.7mm}
{\hspace{-1.5mm}\LARGE\rm
E.~Alecian$^9$ } \\
$^1$ LATT--UMR 5572, CNRS \& Univ.\ P.~Sabatier, 14 Av.\ E.~Belin, F--31400 Toulouse, France \\
$^2$ School of Physics and Astronomy, Univ.\ of St~Andrews, St~Andrews, Scotland KY16 9SS, UK \\
$^3$ LAOG--UMR 5573, CNRS \& Univ.\ J.~Fourier, 31 rue de la Piscine, F--38041 Grenoble, France \\ 
$^4$ School of Physics, Univ.\ of Exeter, Stocker Road, Exeter EX4~4QL, UK \\ 
$^5$ ESO, Karl-Schwarzschild-Str.\ 2, D-85748 Garching, Germany \\ 
$^6$ Department of Physics, Imperial College London, London SW7 2AZ, UK \\ 
$^7$ AAO, PO Box 296, Epping NSW 1710 Australia \\ 
$^8$ CFHT, 65-1238 Mamalahoa Hwy, Kamuela HI, 96743 USA \\ 
$^9$ LESIA--UMR 8109, CNRS \& Univ.\ Paris VII, 5 Place Janssen, F--92195 Meudon Cedex, France 
}

\date{2007, MNRAS, submitted}
\maketitle
 
\begin{abstract}  

From observations collected with the ESPaDOnS and NARVAL spectropolarimeters, 
we report the detection of Zeeman signatures on the classical T~Tauri star (cTTS) 
BP~Tau.  Circular polarisation signatures in photospheric lines and in narrow emission
lines tracing magnetospheric accretion are monitored throughout most of the
rotation cycle of BP~Tau at two different epochs in 2006.  We observe that rotational 
modulation dominates the temporal variations of both unpolarised and circularly 
polarised spectral proxies tracing the photosphere and the footpoints of accretion 
funnels.  

From the complete data sets at each epoch, we reconstruct the large-scale magnetic 
topology and the location of accretion spots at the surface of BP~Tau using 
tomographic imaging.  We find that the field of BP~Tau involves a 1.2~kG dipole and 
1.6~kG octupole, both slightly tilted with respect to the rotation axis.  Accretion 
spots coincide with the two main magnetic poles at high latitudes and overlap with 
dark photospheric  spots;  they cover about 2\% of the stellar surface.  
The strong mainly-axisymmetric poloidal field of BP~Tau is very reminiscent of 
magnetic topologies of fully-convective dwarfs.  It suggests that magnetic fields 
of fully-convective cTTSs such as BP~Tau are likely not fossil remants, but rather 
result from vigorous dynamo action operating within the bulk of their convective zones.  

Preliminary modelling suggests that the magnetosphere of BP~Tau extends to 
distances of at least 4~\rstar\ to ensure that accretion spots are located at high 
latitudes, and is not blown open close to the surface by a putative stellar wind.  
It apparently succeeds in coupling to the accretion disc as far out as the corotation 
radius, and could possibly explain the slow rotation of BP~Tau.  

\end{abstract}


\section{Introduction} 

T~Tauri stars (TTSs) are young low-mass stars that have emerged from their
natal molecular cloud core.   Among them, classical TTSs (cTTSs) are those
still surrounded by accretion discs.  CTTSs host strong magnetic fields 
thought to be responsible for disrupting the central regions of their 
accretion discs and for channelling the disc material towards the
stellar surface along discrete accretion funnels.  This process is 
expected to play a key-role in setting the angular momentum evolution of 
Sun-like protostars \citep[e.g.,][]{Konigl91, Shu94, Cameron93, Romanova04} 
as well as their internal structure.  

One of the prototypical and most extensively studied cTTS is BP~Tau in the 
Taurus star formation region.  BP~Tau hosts multi kG magnetic fields on its 
surface, detected both in photospheric lines \citep[using 
Zeeman broadening, e.g., ][]{Johns99a} and in 
emission lines forming at the base of accretion funnels \citep[using 
Zeeman polarisation, e.g., ][]{Johns99b, Valenti04}.  However, 
the large-scale topology of the magnetic field is not well known, preventing 
detailed studies of how the star magnetically couples to its accretion disc 
and how efficiently angular momentum is transferred from the disc to the 
star \citep[e.g.,][]{Jardine06, Long07}.  

Large-scale topologies of stellar magnetic fields can be constrained 
using Zeeman polarisation signatures in line profiles and the information 
they convey about the field orientation.  By measuring the polarisation 
spectrum of a star and by monitoring its modulation throughout the whole 
rotation cycle, the large-scale topology of the parent surface magnetic 
field can be retreived \citep[e.g.,][]{Donati01b, Donati06b}.  In the particular 
case of cTTSs, this requires Zeeman circular polarisation signatures to be 
detected, not only in emission lines -- tracing the tiny areas  at the 
footpoints of accretion funnels -- but also in photospheric lines -- forming
over most of the stellar surface.  

This technique was recently applied to V2129~Oph, a bright, mildly accreting cTTS 
in the Ophiucus stellar formation region \citep[hereafter D07]{Donati07}.  
The magnetic topology 
of V2129~Oph is found to be significantly more complex than the conventional dipole 
used in most theoretical studies \citep{Romanova03, Romanova04};  despite being 
rather weak (0.35~kG), the dipole field component is nonetheless apparently capable of 
coupling to the accretion disc up to distances of 7 stellar radii.  In particular, 
this result brings further support to the original idea that magnetic coupling 
between the star and its accretion disc is able to control the rotation rate and 
the angular momentum content of the protostar.
In this paper, we apply the same technique to BP~Tau.  

From a detailed spectroscopic study, \citet{Johns99a} conclude that BP~Tau has a 
photospheric temperature \teff\ of $4055\pm112$~K, 
a logarithmic gravity of $3.67\pm0.50$ and a logarithmic metallicity (with 
respect to the Sun) of $0.18\pm0.11$.  Assuming a logarithmic luminosity (with 
respect to the Sun) of $-0.03\pm0.10$ \citep{Gullbring98}, they derive that BP~Tau 
has a radius of $\rstar=1.95\pm0.26$~\rsun.    
The rotation period of BP~Tau, as derived from photometric variability, is reported to 
be $7.6\pm0.1$~d \citep{Vrba86, Shevchenko91};  discrepant estimates, including a 
sudden change of the light-curve period (from 7.6 to 6.1~d, \citealt{Simon90}) are 
also reported, suggesting that the photometric brightness of BP~Tau may sporadically 
include a non-stellar contribution, e.g., from the inner edge of the accretion disc.  
Given the rotation period and the 
line-of-sight projected equatorial rotation velocity \vsini\ \citep[$\simeq10$~\kms\ 
according to][in good agreement with our own estimate of 9~\kms, see below]{Johns99a}, 
we find that $\rstar\sin i\simeq1.4$~\rsun\ and thus that $i\simeq45\degr$.  

Fitting the evolutionary models of \citet{Siess00} to these parameters, we
infer  that BP~Tau is a $0.70\pm0.15$~\msun\ star with an age of about
1.5~Myr.  Models indicate that BP~Tau is still fully convective, making it 
thus significantly different from V2129~Oph (which recently started to build 
up a radiative core, D07) and thus especially interesting for our 
study\footnote{To ease comparison with BP~Tau, we recall that V2129~Oph is 1.35~\msun\ 
star with a radius of 2.4~\rsun, rotating with a period of 6.53~d and accreting 
mass at an estimated rate of about $10^{-8}$~\mspy.  According 
to the models of \citet{Siess00}, V2129~Oph is no longer fully convective 
and hosts a small radiative core weighting about 0.1~\msun.  Further information 
on V2129~Oph can be found in D07.  }.  
Infrared excess and emission lines are conspicuous in the spectrum of BP~Tau, 
indicating the presence of an accretion disc.  Optical veiling, weakening the 
strength of photospheric lines by a significant fraction (typically 10\% to 
50\% depending on epoch and wavelength) also suggests that accretion hot spots 
are present at the surface of the star.  The mass accretion rate of BP~Tau is 
$\Mdot\simeq3\times10^{-8}$~\mspy\ \citep[e.g.,][]{Gullbring98}.  

After presenting and describing our 
spectropolarimetric observations (Secs.~\ref{sec:obs} \& \ref{sec:rot}), we 
apply tomographic imaging on our time-resolved data sets and derive the 
large-scale topology of BP~Tau at 2 different epochs (Sec.~\ref{sec:zdi}) that 
we use to provide a first model of the magnetosphere and accretion funnels 
(Sec.~\ref{sec:mag}).  We finally summarise our results and discuss their 
implications for our understanding of stellar formation (Sec.~\ref{sec:dis}).  

\begin{table*}
\caption[]{Journal of observations.  Columns 1--6 sequentially list the UT date, 
the instrument used, the heliocentric Julian date and UT time (both at mid-exposure), 
and the peak signal to noise ratio (per 2.6~\kms\ velocity bin) of each observation 
(i.e., each sequence of four subexposures).  
Column 7 lists the rms noise level (relative to the unpolarized continuum level 
$I_{\rm c}$ and per 1.8~\kms\ velocity bin) in the circular polarization profile 
produced by Least-Squares Deconvolution (LSD), while column~8 indicates the 
rotational cycle associated with each exposure (within each data set, and using the 
ephemeris given by Eq.~\ref{eq:eph}).  }   
\begin{tabular}{cccccccc}
\hline
Date & Instrument & HJD          & UT      & $t_{\rm exp}$ & \sn\  & $\sigma_{\rm LSD}$ & Cycle \\
(2006) &   & (2,453,000+) & (h:m:s) & (s) &      &   (\ptt)  &  \\
\hline
Feb~07 & ESPaDOnS/CFHT & 773.86074 & 08:38:59 & $4\times600$ & 125 &  5.4 & 0.508 \\ 
Feb~08 & ESPaDOnS/CFHT & 774.72499 & 05:23:37 & $4\times600$ & 130 &  4.7 & 0.622 \\ 
Feb~09 & ESPaDOnS/CFHT & 775.72732 & 05:27:06 & $4\times400$ &  70 & 11.0 & 0.754 \\ 
Feb~10 & ESPaDOnS/CFHT & 776.72808 & 05:28:20 & $4\times500$ & 120 &  5.2 & 0.885 \\ 
Feb~11 & ESPaDOnS/CFHT & 777.72195 & 05:19:38 & $4\times600$ & 120 &  5.5 & 1.016 \\ 
Feb~12 & ESPaDOnS/CFHT & 778.72303 & 05:21:20 & $4\times600$ & 120 &  5.5 & 1.148 \\ 
Feb~13 & ESPaDOnS/CFHT & 779.72259 & 05:20:50 & $4\times600$ &  90 &  8.5 & 1.279 \\ 
Feb~14 & ESPaDOnS/CFHT & 780.72330 & 05:21:59 & $4\times600$ &  70 & 12.6 & 1.411 \\ 
Feb~15 & ESPaDOnS/CFHT & 781.74836 & 05:58:13 & $4\times600$ & 130 &  5.4 & 1.546 \\ 
\hline
Nov~29 & NARVAL/TBL    & 1069.47610 & 23:18:19 & $4\times900$ &  85 &  7.3 & 0.405 \\ 
Nov~30 & ESPaDOnS/CFHT & 1069.98455 & 11:30:31 & $4\times900$ & 140 &  3.5 & 0.472 \\ 
Dec~05 & ESPaDOnS/CFHT & 1074.99539 & 11:46:20 & $4\times900$ & 170 &  2.7 & 1.131 \\ 
Dec~06 & ESPaDOnS/CFHT & 1075.97733 & 11:20:23 & $4\times900$ & 170 &  2.8 & 1.260 \\ 
Dec~07 & ESPaDOnS/CFHT & 1076.90044 & 09:29:42 & $4\times900$ & 170 &  2.9 & 1.382 \\ 
Dec~08 & ESPaDOnS/CFHT & 1077.86802 & 08:43:04 & $4\times800$ & 170 &  2.8 & 1.509 \\ 
Dec~09 & ESPaDOnS/CFHT & 1078.78979 & 06:50:29 & $4\times600$ & 120 &  4.2 & 1.630 \\ 
Dec~10 & ESPaDOnS/CFHT & 1079.79637 & 07:00:00 & $4\times600$ &  80 &  6.6 & 1.763 \\ 
\hline
\end{tabular}
\label{tab:log}
\end{table*}

\section{Observations}
\label{sec:obs}

\subsection{Spectropolarimetric data}

Spectropolarimetric observations of BP~Tau were collected in 2006 at 2 different 
epochs, using mostly ESPaDOnS on the 3.6~m Canada-France-Hawaii Telescope (CFHT) 
in Hawaii and sporadically NARVAL on the 2~m T\'elescope Bernard Lyot (TBL) in 
southern France.  ESPaDOnS and NARVAL are twin instruments able to collect 
stellar spectra spanning the whole spectral domain (from 370 to 1,000~nm) at 
a resolving power of 65,000, in either circular or linear polarisation 
\citep[][Donati et al., in preparation]{Donati06c}.  A total of 17 circular 
polarisation spectra were collected in 2006 February (9 ESPaDOnS spectra) and 
November/December (7 ESPaDOnS and 1 NARVAL spectra); both data sets are referred 
to as Feb06 and Dec06 in the following.  Each polarisation spectrum consists of 
4 individual subexposures taken in different polarimeter configurations to remove 
all spurious polarisation signatures at the first order.   

Raw frames are processed with {\sc Libre~ESpRIT}, a fully automatic reduction 
package/pipeline developed and owned by JFD and installed both at CFHT and TBL.  
It automatically performs optimal extraction of ESPaDOnS and NARVAL unpolarised 
(Stokes $I$) and circularly polarised (Stokes $V$) spectra following the 
procedure described in \citet[][Donati et al., in preparation]{Donati97}.   
The velocity step corresponding to CCD pixels is about 2.6~\kms;  however, thanks 
to the fact that the spectrograph slit is tilted with respect to the CCD lines, 
spectra corresponding to different CCD columns across each order feature a 
different pixel sampling.  {\sc Libre~ESpRIT} uses this opportunity to carry out 
optimal extraction of each spectrum on a sampling grid denser than the original 
CCD sampling, with a spectral velocity step set to about 0.7 CCD pixel 
(i.e.\ 1.8~\kms).  
The peak signal-to-noise ratios (per 2.6~\kms\ velocity bin) achieved on the 
collected spectra (i.e., the sequence of 4 subexposures) range between 70 and 
170 depending on the weather and the exposure time.  
The full journal of observations is presented in Table~\ref{tab:log}.   

\begin{figure}
\center{\includegraphics[scale=0.32,angle=-90]{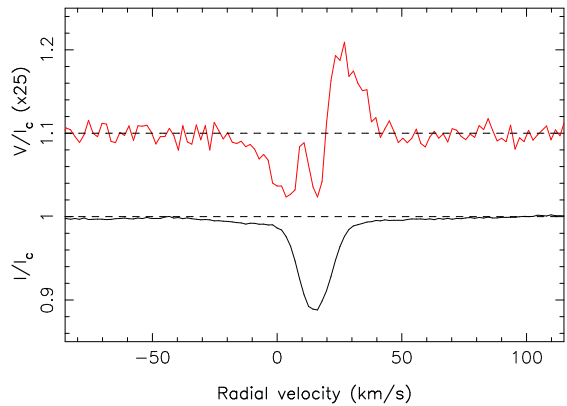}} 
\caption[]{LSD circularly-polarized and unpolarized
profiles of BP~Tau (top, bottom curves respectively) on 2006 Feb~11 (phase 0.016).  
The mean polarization profile is expanded by a factor of 25 and shifted upwards 
by 1.1 for display purposes.  }
\label{fig:lsd}
\end{figure}

\begin{figure*}
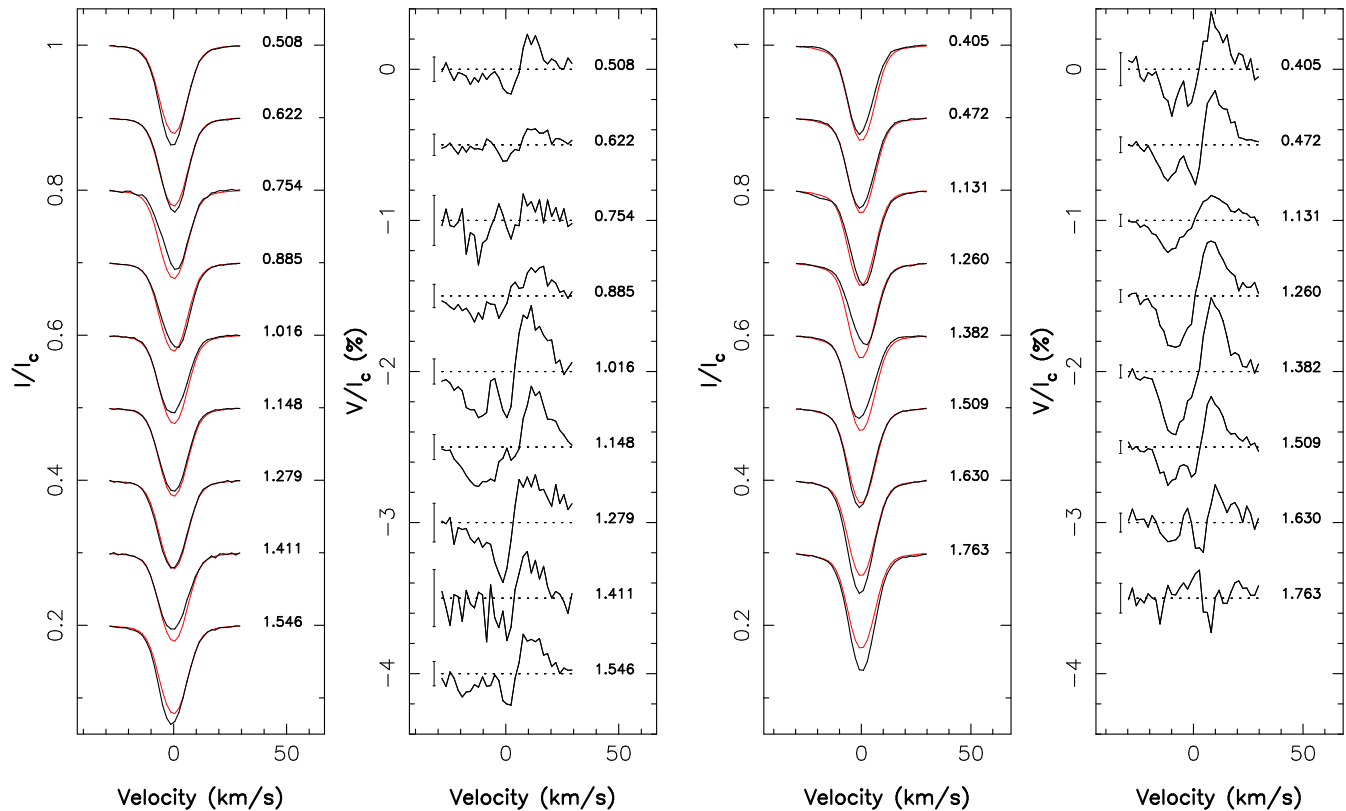

\center{\hbox{\includegraphics[scale=0.58,angle=-90]{fig/bptau_poli1.ps}\hspace{1mm}
              \includegraphics[scale=0.58,angle=-90]{fig/bptau_polv1.ps}\hspace{4mm}
              \includegraphics[scale=0.58,angle=-90]{fig/bptau_poli2.ps}\hspace{1mm}
              \includegraphics[scale=0.58,angle=-90]{fig/bptau_polv2.ps}}}
\caption[]{Stokes $I$ (panels 1 \& 3) and Stokes $V$ LSD profiles (panel 2 \& 4)
of photospheric lines of BP~Tau (thick black line), for each observing night (top to bottom) 
of both Feb06 (left) and Dec06 (right) runs.  The mean Stokes $I$ LSD profile (averaged over 
the full set within each run, thin red line) is also shown to emphasise temporal variations.  
The rotational cycle associated with each observation is noted next to each profile;  
3$\sigma$ error bars are also shown to the left of each Stokes $V$ profile.
All profiles are plotted in the velocity rest frame of BP~Tau.  }
\label{fig:pho}
\end{figure*}

Least-Squares Deconvolution (LSD; \citealt{Donati97}) was applied to all
observations.   The line list we employed for LSD is computed from an {\sc
Atlas9} LTE model atmosphere  \citep{Kurucz93} and corresponds to a K7IV
spectral type ($\teff=4,000$~K and  $\logg=3.5$) appropriate for BP~Tau.  
We selected only  moderate to strong spectral lines whose synthetic 
profiles had line-to-continuum core depressions larger than 40\% neglecting 
all non-thermal broadening mechanisms.  We omitted the spectral regions within 
strong lines not formed mostly in the photosphere, such as the Balmer and He 
lines, and the \caii\ H, K and infrared triplet (IRT) lines. 
Altogether, about 9,400 spectral features are used in this process, most of
them from \fei.  Expressed in units of the unpolarised continuum level 
$I_{\rm c}$, the average noise levels of the  resulting LSD signatures range 
from 2.7 to 12.6$\times10^{-4}$  per 1.8~\kms\ velocity bin.  

All relevant information is summarised in Table~\ref{tab:log}.  
Rotational cycles $E$ are computed from heliocentric Julian dates according 
to the ephemeris:  
\begin{equation}
\mbox{HJD} = 2453770.0 + 7.6 E. 
\label{eq:eph}
\end{equation}
Note that the rotation cycle of BP~Tau (7.6~d) was 
fully covered in Feb06, but only about 75\% of it in Dec06.  The error on the 
period (0.1~d) translates into a phase uncertainty of 0.5 rotation cycle 
between the Feb06 and Dec06 data sets (separated by 298~d or 39 rotations).  

\subsection{Zeeman signatures}

Zeeman signatures, featuring full amplitudes of about 0.5\% of the unpolarised 
continuum level on average, are clearly detected in the LSD profiles of all spectra 
(e.g., see Fig.~\ref{fig:lsd}).  The complete sets of LSD Stokes $I$ and $V$ 
profiles for both Feb06 and Dec06 runs are shown in Fig.~\ref{fig:pho}.  
Note in particular that Zeeman signatures are significantly wider than unpolarised 
profiles, suggesting that field strengths at the surface of BP~Tau are rather 
strong.    
The corresponding longitudinal fields \citep[computed from the first order moment
of the Stokes $V$ profiles,][]{Donati97} are typically equal to a few hundred G;  
while they clearly vary with time, they keep the same (negative) sign throughout 
both runs (see Table~\ref{tab:beff}).  

We divided the LSD line list into 2 subsets, a blue and a red subset including 
lines bluer and redder than 620~nm respectively;  from those partial line lists, 
we derived LSD signatures (not shown) and corresponding longitudinal fields.  We find that 
the longitudinal fields derived from the blue and red line subsets are slightly 
(about 15\%) larger and smaller respectively than the average longitudinal fields 
derived from the original LSD signatures.  This is similar (though much less 
extreme) than what was reported on the cool dwarf $\xi$~Boo \citep{Petit05};  
the origin of this effect is not clear.  

The longitudinal fields we derive are significantly larger than the upper limits 
obtained by \citet{Johns99b} and \citet{Valenti04}.  This discrepancy is due 
to the different methods used to estimate longitudinal fields;  the method 
used here (equivalent to the well-known ``centre of gravity'' technique) is known 
to be very robust for all field configurations and strengths \citep{Landi04}, while 
cross-correlation methods (used by \citealt{Johns99b}) are found to yield 
underestimates of both the longitudinal field and the corresponding error bar when 
Zeeman signatures are significantly broader than the unpolarised profile and 
feature a weak amplitude (as in the present case);  
long-term intrinsic variability of the large-scale magnetic topology may also 
partly explain the discrepancy.  

We also estimate the amount of veiling in the spectrum of BP~Tau (at an average 
wavelength of 620~nm) by comparing the 
equivalent widths of Stokes $I$ LSD photospheric profiles between BP~Tau and a standard 
star of similar spectral type ($\delta$~Eri), using the same line list for both stars;  
the veiling parameter we derive varies between 0 and 60\% over the full span of our 
observations, i.e., comparable to values previously published in the literature 
\citep{Gullbring98};  the relative veiling variation over one single run is about 
$\pm15$\%, similar to previous reports by \citet{Valenti03}.  

\begin{table*}
\caption[]{Longitudinal magnetic field of BP~Tau and associated error bars, as estimated 
from the LSD profiles ($B_{\rm LSD}$, col.~2), 
the \caii\ IRT emission core ($B_{\rm IRT}$, col.~3),  
the \hei\ 587.562~nm ($D_3$) line ($B_{\rm \hei\ D_3}$, col.~4), 
the \hei\ 667.815~nm line ($B_{\rm \hei\ 667}$, col.~5) and 
the 3 \feii\ lines of multiplet 32 around 500~nm ($B_{\rm \feii}$, col.~6), \hal\ ($B_{\rm \hal}$, col.~7) and \hbe\ 
($B_{\rm \hbe}$, col.~8).  
Column 9 lists the veiling parameter $r$ at each phase at an average wavelength of 620~nm  
(with typical error bars of about 0.01), defined as the relative 
difference in the equivalent width of Stokes $I$ LSD photospheric profiles 
between BP~Tau and the standard star $\delta$~Eri.  
Rotational cycles (col.~1) are computed according to Eq.~\ref{eq:eph}. }
\begin{tabular}{ccccccccc}
\hline
Cycle  & $B_{\rm LSD}$ & $B_{\rm IRT}$ & $B_{\rm \hei\ D_3}$ & $B_{\rm \hei\ 667}$  & $B_{\rm \feii}$ & $B_{\rm \hal}$  & $B_{\rm \hbe}$ & $r$ \\
       & (kG) & (kG) & (kG) & (kG) & (kG) & (kG) & (kG) & \\
\hline 
\multicolumn{9}{c}{Feb06 run} \\ 
\hline 
0.508 & $-0.19\pm0.05$ & $0.53\pm0.07$ & $1.77\pm0.23$ & $5.32\pm0.66$ & $2.84\pm1.10$ & $0.01\pm0.02$ & $0.61\pm0.17$ & 0.22 \\
0.622 & $-0.14\pm0.05$ & $0.44\pm0.07$ & $1.44\pm0.18$ & $1.74\pm0.44$ & $1.64\pm0.61$ & $0.00\pm0.02$ & $0.43\pm0.11$ & 0.29 \\
0.754 & $-0.21\pm0.15$ & $0.64\pm0.12$ & $1.39\pm0.28$ & $3.93\pm0.50$ & $1.92\pm0.63$ & $0.02\pm0.03$ & $0.79\pm0.34$ & 0.59 \\
0.885 & $-0.33\pm0.07$ & $1.09\pm0.07$ & $2.34\pm0.15$ & $4.36\pm0.33$ & $3.67\pm0.95$ & $0.08\pm0.02$ & $0.60\pm0.12$ & 0.35 \\
1.016 & $-0.69\pm0.14$ & $1.57\pm0.08$ & $3.30\pm0.12$ & $5.58\pm0.25$ & $4.50\pm0.86$ & $0.23\pm0.02$ & $1.36\pm0.15$ & 0.51 \\
1.148 & $-0.63\pm0.13$ & $1.56\pm0.09$ & $3.75\pm0.17$ & $8.17\pm0.47$ & $4.40\pm1.02$ & $0.12\pm0.02$ & $1.05\pm0.14$ & 0.42 \\
1.279 & $-0.50\pm0.16$ & $1.23\pm0.12$ & $3.43\pm0.31$ & $7.88\pm0.88$ & $3.62\pm1.07$ & $0.14\pm0.03$ & $0.91\pm0.25$ & 0.32 \\
1.411 & $-0.19\pm0.15$ & $0.66\pm0.14$ & $1.68\pm0.55$ & $7.45\pm1.45$ & $3.93\pm1.67$ & $0.06\pm0.03$ & $1.06\pm0.41$ & 0.48 \\
1.546 & $-0.31\pm0.07$ & $0.74\pm0.07$ & $1.85\pm0.19$ & $4.80\pm0.54$ & $2.67\pm0.91$ & $0.04\pm0.02$ & $0.36\pm0.18$ & 0.21 \\
\hline
\multicolumn{9}{c}{Dec06 run} \\ 
\hline
0.405 & $-0.29\pm0.08$ & $0.75\pm0.08$ & $1.73\pm0.22$ & $3.96\pm0.43$ & $1.57\pm0.46$ & $0.06\pm0.02$ & $0.15\pm0.14$ & 0.39 \\
0.472 & $-0.37\pm0.04$ & $0.92\pm0.05$ & $2.36\pm0.12$ & $4.38\pm0.29$ & $2.64\pm0.58$ & $0.05\pm0.02$ & $0.80\pm0.09$ & 0.31 \\
1.131 & $-0.28\pm0.03$ & $0.64\pm0.04$ & $2.02\pm0.13$ & $4.95\pm0.35$ & $2.19\pm0.46$ & $0.02\pm0.02$ & $0.42\pm0.07$ & 0.20 \\
1.260 & $-0.53\pm0.05$ & $0.83\pm0.04$ & $2.37\pm0.11$ & $3.71\pm0.22$ & $2.01\pm0.36$ & $0.04\pm0.02$ & $0.40\pm0.07$ & 0.39 \\
1.382 & $-0.60\pm0.04$ & $0.72\pm0.04$ & $1.85\pm0.09$ & $3.31\pm0.18$ & $2.64\pm0.49$ & $0.06\pm0.02$ & $0.52\pm0.08$ & 0.41 \\
1.509 & $-0.30\pm0.03$ & $0.93\pm0.05$ & $2.62\pm0.12$ & $4.95\pm0.30$ & $3.10\pm0.81$ & $0.05\pm0.02$ & $0.93\pm0.09$ & 0.20 \\
1.630 & $-0.12\pm0.03$ & $1.17\pm0.07$ & $3.70\pm0.22$ & $7.78\pm0.72$ & $3.67\pm1.00$ & $0.07\pm0.02$ & $0.46\pm0.12$ & 0.06 \\
1.763 & $-0.06\pm0.05$ & $0.83\pm0.11$ & $2.28\pm0.38$ & $5.82\pm1.34$ & $3.09\pm1.21$ & $0.07\pm0.03$ & $0.20\pm0.24$ & 0.00 \\
\hline
\end{tabular}
\label{tab:beff}
\end{table*}

Circular polarisation is also detected in most emission lines, and in particular 
in the \hei\ 587.562~nm ($D_3$) line and in the \caii\ IRT lines known as good  tracers of 
magnetospheric accretion \citep{Johns99b, Valenti04} with medium magnetic sensitivity 
(Land\'e factors of 1.0);  clear signatures are also detected at all times in the 
\hei\ 667.815~nm line (Land\'e factor of 1.0) and in the \feii\ 492.393~nm, 501.842~nm and 
516.903~nm lines (multiplet 32, average Land\'e factor of 1.7).   The complete 
sets of Stokes $I$ and $V$ profiles corresponding to the \caii\ 
IRT\footnote{Note that the 3 components of the \caii\ IRT were averaged together 
in a single profile to increase \sn\ further.} and the \hei\ $D_3$ emission lines 
are shown in Figs.~\ref{fig:irt} and \ref{fig:he} respectively.   The longitudinal 
fields (see Table~\ref{tab:beff}) are significantly larger, and of opposite sign, 
than those derived from photospheric lines.  They reach up to 1.6~kG and 3.8~kG for the 
\caii\ IRT and the \hei\ $D_3$ emission lines respectively, and up to 8.2~kG for 
the \hei\ 667.815~nm line;  longitudinal fields from the \feii\ lines more or less 
repeat (with larger error bars) those traced by the \hei\ $D_3$ line.  
Note that the field values 
we derive from the \hei\ $D_3$ line are similar to those measured by 
\citet{Johns99a}, \citet{Valenti04} and \citet{Symington05}.

\section{Rotational modulation}
\label{sec:rot}

We now examine the temporal variations of the Stokes $I$ and $V$ profiles of 
photospheric and emission lines throughout the whole rotation cycle of BP~Tau and 
demonstrate that these variations can be assigned to rotational modulation.  

\subsection{Photospheric lines and accretion proxies}

As for V2129~Oph (D07), the emission lines tracing accretion regions at the surface of 
the star (and in particular the \caii\ IRT and the \hei\ 587.562~nm and 667.815~nm lines) 
exhibit the strongest and simplest evolution over the timescale of our observations.  
The level of \hei\ emission (e.g., Figs.~\ref{fig:he} \& \ref{fig:ewblonvrad}) undergoes a progressive increase 
followed by a regular decrease over the rotation period, i.e., very similar to what is 
reported by \citet{Valenti03}.  Moreover, comparing, e.g., \hei\ profiles at phases 0.508 
and 1.546 on the Feb06 run demonstrates that the amount of emission more or less returns 
to its initial level once the star completed a complete rotation cycle (see Fig.~\ref{fig:ewblonvrad});  
the main part 
of the observed variability may thus be safely attributed to rotational modulation.   

\begin{figure*}
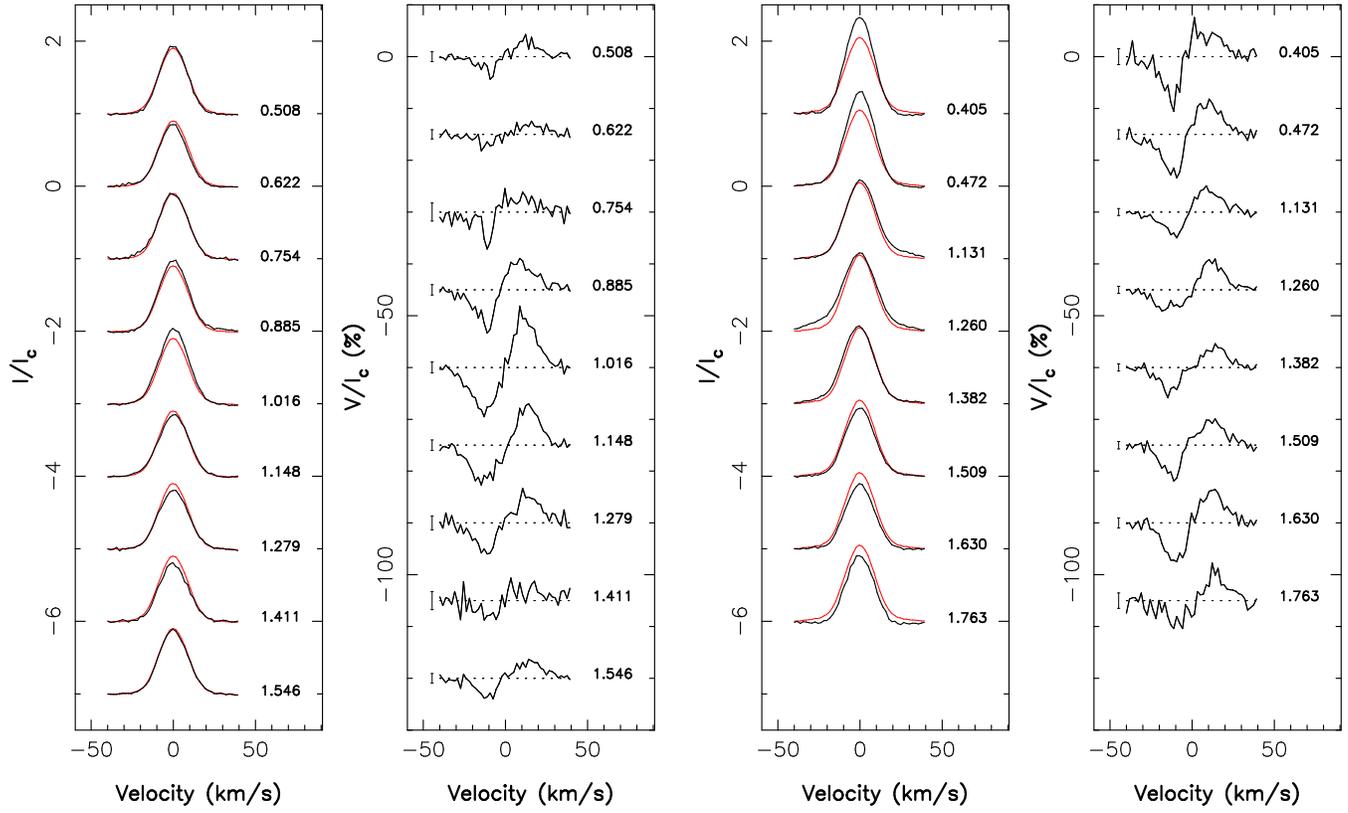

\center{\hbox{\includegraphics[scale=0.58,angle=-90]{fig/bptau_irti1.ps}\hspace{1mm}
              \includegraphics[scale=0.58,angle=-90]{fig/bptau_irtv1.ps}\hspace{4mm}
              \includegraphics[scale=0.58,angle=-90]{fig/bptau_irti2.ps}\hspace{1mm}
              \includegraphics[scale=0.58,angle=-90]{fig/bptau_irtv2.ps}}}
\caption[]{The same as Fig.~\ref{fig:pho} but for the \caii\ IRT lines.  } 
\label{fig:irt}
\end{figure*}

\begin{figure*}
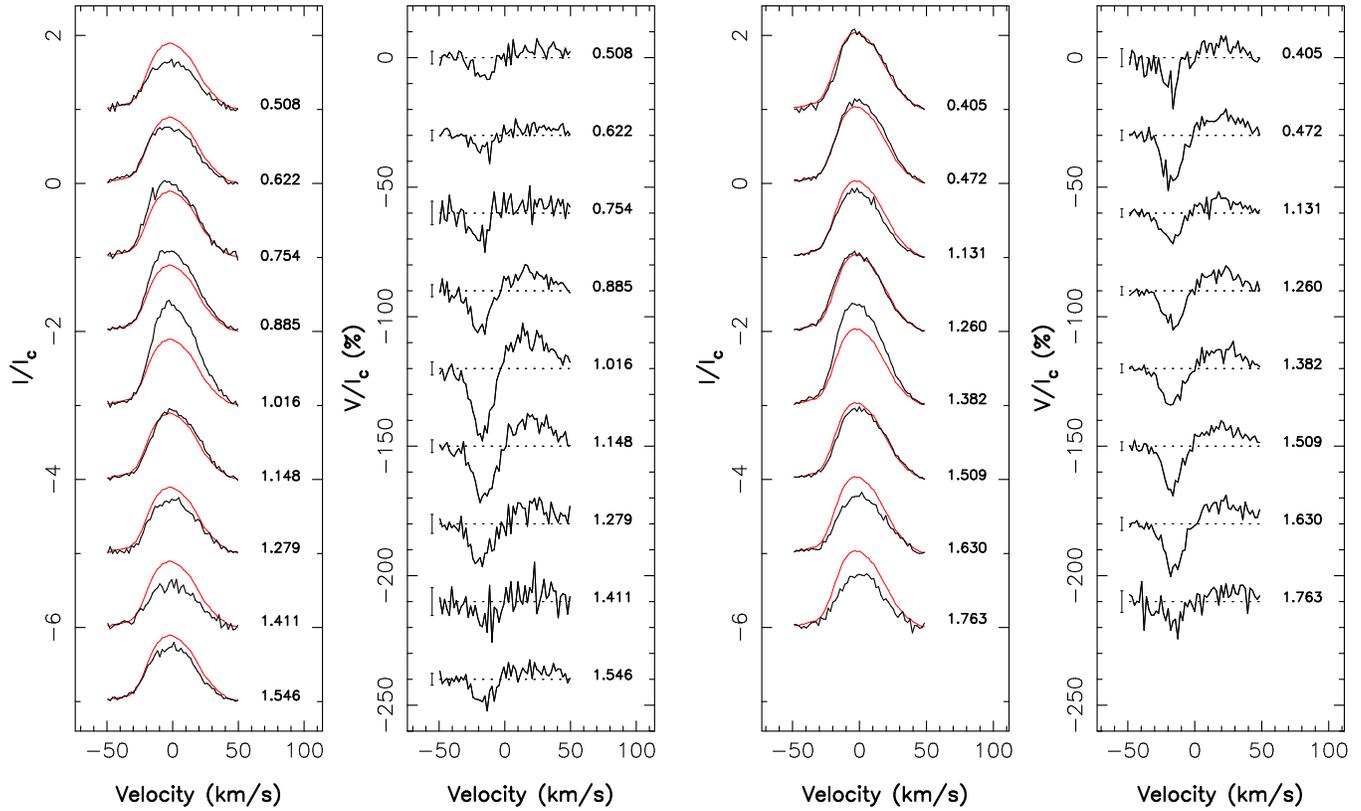

\center{\hbox{\includegraphics[scale=0.58,angle=-90]{fig/bptau_hei1.ps}\hspace{1mm}
              \includegraphics[scale=0.58,angle=-90]{fig/bptau_hev1.ps}\hspace{4mm}
              \includegraphics[scale=0.58,angle=-90]{fig/bptau_hei2.ps}\hspace{1mm}
              \includegraphics[scale=0.58,angle=-90]{fig/bptau_hev2.ps}}}
\caption[]{The same as Fig.~\ref{fig:pho} but for the \hei\ 587.562~nm ($D_3$) line.  } 
\label{fig:he}
\end{figure*}

Looking at the variability of the \caii\ IRT emission profiles in Feb06 brings the same 
conclusion (see Fig.~\ref{fig:irt}), eventhough the amount of fluctuation -- $\pm10$\% about 
the mean (i.e., from 18 to 22~\kms\ or 0.051 to 0.062~nm) -- is smaller than that of the \hei\ 
$D_3$ line -- almost a factor of 2 peak to peak (i.e., from 30 to 60~\kms\ or 0.06 to 0.12~nm).  
The variation of the \hei\ 667.815~nm line is even more extreme, reaching up to a factor of 
3 peak to peak (from 6 to 18~\kms\ or 0.013 to 0.040~nm);  the \feii\ emission lines  
vary in strength by about a factor of 2 (from 7 to 14~\kms\ or 0.012 to 0.024~nm).  
Note however that the \caii\ IRT emission profiles at phases 0.405 and 0.472 in the Dec06 
run (both taken about 1 rotation cycle earlier than the bulk of the other Dec06 data) are 
significantly stronger than the average, and in particular stronger than what they should have 
been in case of pure rotational modulation (given the observed profiles at phases 1.382 and 
1.509).  We attribute this to intrinsic variability\footnote{Other emission lines are 
mostly unaffected by this intrinsic variability.}, which we corrected by scaling down both 
profiles by 20\% (Figs.~\ref{fig:irt} and \ref{fig:ewblonvrad} show profiles/equivalent widths prior 
to this correction).  
Maximum emission is reached (in all lines simultaneously) at phases 1.02 and 1.38 in the 
Feb06 and Dec06 runs respectively (see Fig.~\ref{fig:ewblonvrad}).  

Stokes $V$ profiles from emission lines strengthen the evidence that rotational 
modulation dominates the observed temporal variations.  The Feb06 data are particularly 
clear in this respect;  the shape of Zeeman signatures is indeed varying very smoothly 
with rotation phase (reaching maximum strength around phase 1.02) and repeats well 
after one complete rotation cycle.  The corresponding longitudinal field variations with 
rotational cycle, shown in Fig.~\ref{fig:ewblonvrad}, further confirm this impression.  
The Dec06 Zeeman data also show definite (though smaller) variability with rotation 
phase.  The overall similitude between both data sets (with a smooth rise and decrease 
of the line emission and Zeeman signature amplitude) throughout the rotation period 
suggests that both line emission and Zeeman signatures were likely weak in the 
Dec06 phase gap during which no observation were collected;  we thus conclude that 
all the large magnetic regions have most likely been observed in Dec06 despite the 
incomplete phase coverage.  

The longitudinal fields we measure, although grossly similar at both observing epochs, 
nevertheless exhibit a different rotational modulation in Feb06 and Dec06 
(see Fig.~\ref{fig:ewblonvrad});  for instance, the longitudinal field curve shows only 
one maximum in Feb06, while it shows two distinct peaks in Dec06.  It suggests 
that intrinsic variability likely distorted the large-scale field topology between both 
observing epochs.  
Moreover, at both epochs, the longitudinal field curves are not symmetric with respect 
to the phase of maximum line emission, suggesting that the field within accretion spots 
is not purely radial but rather slightly tilted with respect to the local meridional plane.  
In Feb06 for instance, the phases at which maximum longitudinal fields are observed are 
shifted with respect to the phase of maximum line emission (1.02) and differ for each 
line, with \hei\ lines (and in particular the \hei\ 667.815~nm line) showing maximum 
phase shifts;  it suggests that the tilt (with respect to the local meridional plane) 
of the magnetic field within accretion spots is larger in the formation region of \hei\ 
lines.  

\begin{figure*}
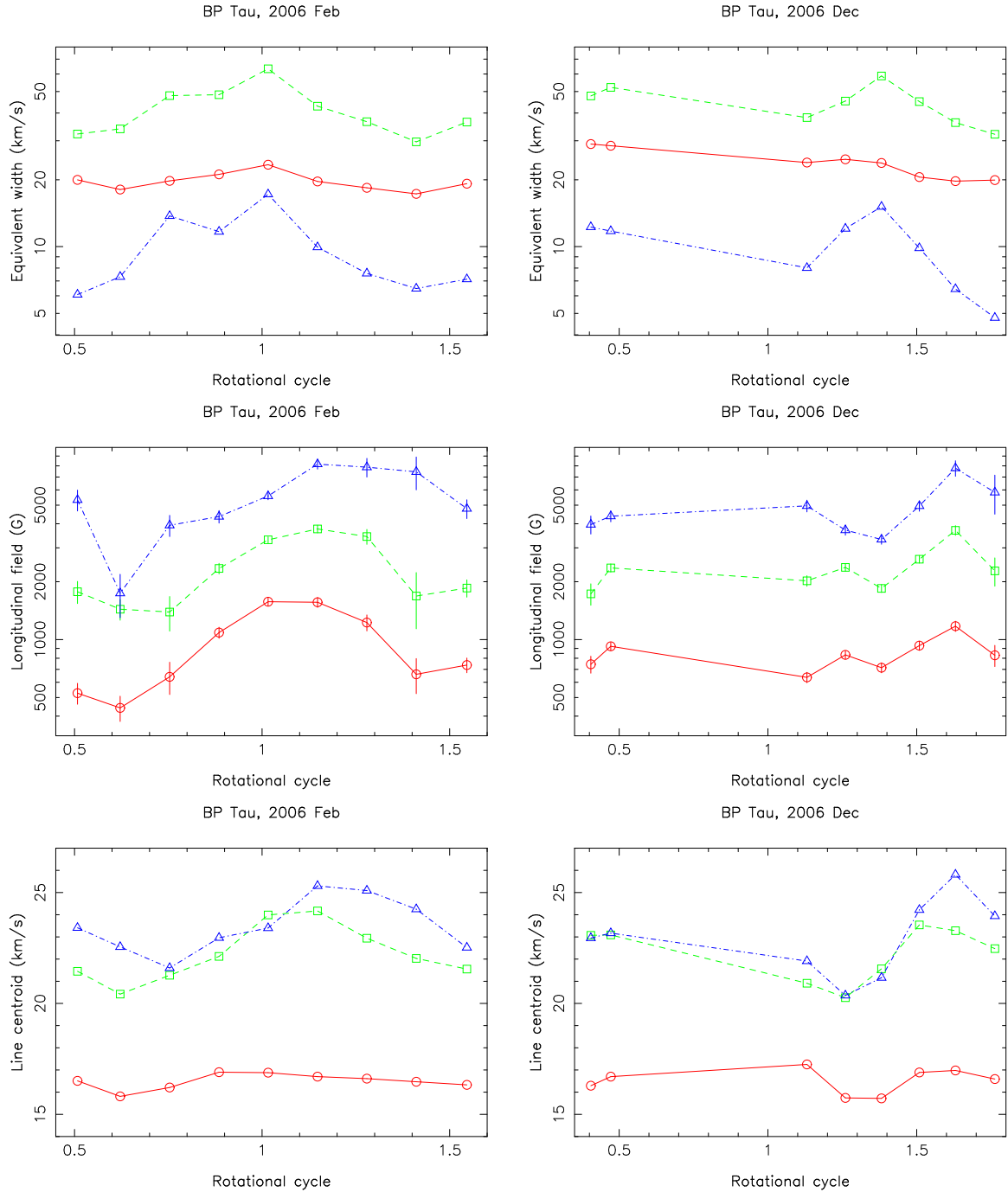

\center{\includegraphics[scale=0.32,angle=-90]{fig/bptau_ew1.ps}\hspace{5mm} 
\includegraphics[scale=0.32,angle=-90]{fig/bptau_ew2.ps}} 
\center{\includegraphics[scale=0.32,angle=-90]{fig/bptau_beff1.ps}\hspace{5mm} 
\includegraphics[scale=0.32,angle=-90]{fig/bptau_beff2.ps}} 
\center{\includegraphics[scale=0.32,angle=-90]{fig/bptau_vrad1.ps}\hspace{5mm} 
\includegraphics[scale=0.32,angle=-90]{fig/bptau_vrad2.ps}} 
\caption[]{Equivalent widths (top), longitudinal fields (with $\pm$1$\sigma$ error bars, 
centre) and line centroids (bottom, the stellar velocity rest frame being at 15.5~\kms) 
as derived from the \caii\ IRT (red circles), the \hei\ $D_3$ (green squares) and 
the \hei\ 667.815~nm (blue triangles) emission lines as a function 
of rotational cycle, for both Feb06 (left) and Dec06 (right) runs.    
Relative error bars on equivalent widths (top panels) are smaller than the symbol 
size.  Note the logarithmic vertical scale for the top and centre panels.  }
\label{fig:ewblonvrad}
\end{figure*}

The Stokes $V$ LSD photospheric profiles also show clear temporal variations 
(see  Fig.~\ref{fig:pho}).  Zeeman signatures in Feb06, tracing negative longitudinal fields 
throughout the whole rotation cycle, vary smoothly from small (e.g., phase 0.622) to large 
(e.g., phase 1.016) amplitudes, then decrease again to their initial sizes and shapes 
(e.g., phase 1.546 and 0.508).  Similar behaviour is observed on Dec06 with maxium and 
minimum Zeeman signals around phase 1.38 and 1.76 respectively.  This variability  
strongly argues in favour of rotational modulation.  Zeeman signatures are 
more complex than their emission line counterparts and trace a field of opposite polarity;  
it suggests that emission and photospheric lines do not form over the same regions of the 
stellar surface.  A similar conclusion was reached in the case of V2129~Oph (D07).  

The Stokes $I$ LSD profiles display only modest variability, apart from the changes 
in equivalent widths ($\pm15$\% peak to peak within each run, up to 60\% over both runs) 
that we attribute to veiling (see
Table~\ref{tab:beff}).  We find that statistically, veiling tends to be larger when 
emission lines are stronger;  this correlation is however rather loose (as already 
pointed at by \citealt{Valenti03}), with strong veiling episodes sometimes occuring 
when emission lines are weak (e.g., at phase 1.411 in Feb06).  Apart from veiling, 
variations in Stokes $I$ photospheric profiles are also visible (though modest) and 
are reminiscent of bump-like signatures from dark spots travelling from the blue 
to the red side of the line profile (e.g., from phase 0.754 to phase 1.016 in Feb06, 
or from phase 1.260 to 1.382 in Dec06);  at this point, and given the low \vsini\ 
of BP~Tau, this assumption is no more than a speculation and surface imaging tools 
are needed to test it in more details.  

The average radial velocity of the photospheric lines is 15.7~\kms\ in Feb06 
and 15.3~\kms\ in Dec06, suggesting an average radial velocity of about 15.5~\kms;  
typical variations of up to $\pm0.8$~\kms\ are observed throughout each run.  
We assume in the following that this average position represents the heliocentric 
radial velocity of the stellar rest frame.   The average width of photospheric lines 
(full width at half maximum of 15~\kms) is mostly due to rotation ($\vsini=9$~\kms).  

The \caii\ IRT emission core is centred at 16.5~\kms\ on average, i.e., redshifted 
by about 1~\kms\ with respect to the photospheric lines.  As for photospheric lines, 
variations of up to $\pm0.8$~\kms\ are observed throughout each run, with a smooth 
dependance with rotational phase (see Fig.~\ref{fig:ewblonvrad}).  
The width of the central emission core (full width at half 
maximum of 21~\kms) is comparable to the rotational broadening of the star.  
The \feii\ emission lines are centred at 17.3~\kms\ on average and feature a full 
width at half maximum of about 24~\kms.  Velocity variations are stronger, reaching 
up to $\pm2$~\kms throughout both runs.  

Both \hei\ emission lines exhibit higher redshifts relative to the stellar 
rest frame than the \caii\ IRT emission core and lie at 23~\kms\ on average, i.e., 
about 7.5~\kms\ redward of the photospheric lines.  Larger variations of the line 
centroids (of $\pm2-3$~\kms) are also observed throughout each run;  the phase 
dependance is rather smooth and repeats well between both lines, suggesting that it 
is caused by rotational modulation (see Fig.~\ref{fig:ewblonvrad}).  Since 
maximum \hei\ emission occurs roughly half-way through the blue to red migration 
of the \hei\ emission lines (i.e., at phase 1.00 and 1.40 for the Feb06 and Dec06 runs 
respectively), we speculate that this is likely where the parent accretion spots are 
located.  While the \hei\ 
667.815~nm line is roughly as wide as the \caii\ emission cores (full width at half 
maximum of 22~\kms), the \hei\ $D_3$ line is twice as wide (full width at half 
maximum of 44~\kms) as a result of being a composite profile of 6 different transitions.  

A broad \caii\ emission component (with a full-width at half maximum of about 200~\kms, 
similar to what shown by, e.g., \citealt{Gullbring96b} in their Fig.~6) is also present 
(though not shown here) in some of our spectra, mostly those from Dec06.  Its red wing 
more or less anticorrelates with the narrow central component, featuring an 
emission bump around phases of minimum core emission (phase 1.63 in Dec06) and a relative 
dip (with respect to the mean profile) around phases of maximum core emission (phase 
1.26--1.38 in Dec06).  As it correlates well with the red wing of Balmer lines, we 
attribute it to disc material free-falling along accretion funnels (see below).  
This broad component is apparently subject to a higher level of intrinsic 
variability than the narrow component.  
Stokes $V$ signatures being entirely confined to the central narrow emission component, 
we do not consider the broad component any further in the present study and simply 
removed it from all \caii\ profiles (and in particular from those shown in Fig.~\ref{fig:irt}).  
No similar broad emission component is visible in \hei\ and \feii\ lines.  

The widths and redshifts of the \caii, \feii\ and \hei\ Zeeman signatures are similar to 
those of the unpolarised emission profiles.  However, while the Stokes $V$ signatures 
from the \caii\ and \feii\ emission lines are roughly antisymmetric (with respect to 
the line centre), those from \hei\ lines show significant departures from antisymmetry, 
with the blue lobe being both narrower and deeper than the red lobe.  
It suggests that Zeeman signatures from \hei\ lines form in regions featuring 
significant velocity gradients, whereas Zeeman signatures from \caii\ and \feii\ lines form 
in regions that are almost at rest.  This is in good agreement with the observed line 
velocity redshifts, much larger for the \hei\ lines than for the \caii\ and \feii\ lines.  
We also note that both \hei\ lines trace larger toroidal fields than the \caii\ and \feii\ 
lines (see above).  

Most observations reported here are similar to what V2129~Oph exhibits (D07), suggesting 
that the same modelling strategy can be employed.  We thus assume, as for V2129~Oph, 
that emission lines comprise two physically distinct components\footnote{These two 
modelling components should not be confused with the narrow and broad components of 
the \caii\ emission lines mentionned above, and are both used to describe the narrow 
line emission core only.}.   We attribute the first of 
these, the accretion component, to localised accretion spots at the surface of the star 
whose visibility varies as the star rotates, giving rise to rotational modulation of
the emission (both in intensity and radial velocity).  A second, chromospheric component 
is more or less evenly distributed over the surface 
of the star, producing a time-independent emission component.  We further assume that 
the chromospheric component is mostly unpolarised while Stokes $V$ signatures arise only 
in the accretion component.  With the relative strengths of both components varying from 
line to line, we expect different lines to yield different longitudinal fields;  moreover, 
we expect lines showing higher levels of variability to yield stronger longitudinal fields 
and larger velocity variations, in agreement with what we observe on BP~Tau.  
This model is further detailed in Sec.~\ref{sec:zdi}.  

\begin{figure*}
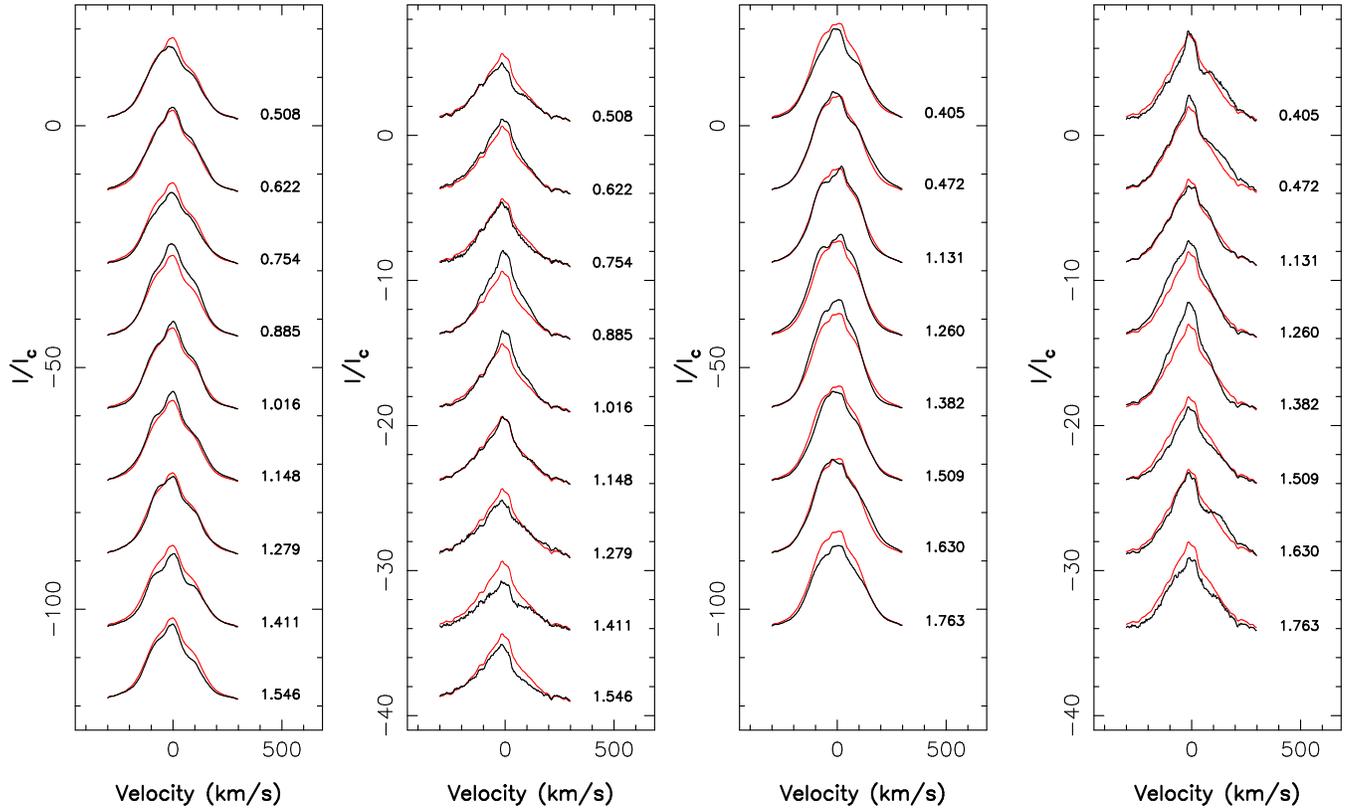

\center{\hbox{\includegraphics[scale=0.58,angle=-90]{fig/bptau_hal1.ps}\hspace{1mm}
              \includegraphics[scale=0.58,angle=-90]{fig/bptau_hbe1.ps}\hspace{1mm}
              \includegraphics[scale=0.58,angle=-90]{fig/bptau_hal2.ps}\hspace{4mm}
              \includegraphics[scale=0.58,angle=-90]{fig/bptau_hbe2.ps}}}
\caption[]{Temporal variations of the H$\alpha$ (panels 1 \& 3) and H$\beta$ (panels 2 
\& 4) profiles of BP~Tau (thick black line), for each observing night (top to bottom) 
of both Feb06 (left) and Dec06 (right) runs.  The mean profile (averaged over the full set 
within each run, thin red line) is also shown to emphasise temporal variations.  
The rotational cycle associated with each observation is noted next to each profile.  
All profiles are plotted in the velocity rest frame of BP~Tau.  }
\label{fig:bal}
\end{figure*}

\subsection{Balmer lines}

H$\alpha$ and H$\beta$ lines exhibit strong emission with average
equivalent widths of 4,850 and 1,210~\kms, (i.e., 10.6 and 2.0~nm) respectively.
Maximum emission is reached at phase 0.89 and 1.38 in the Feb06 and Dec06 runs 
respectively, i.e., at roughly the same time as all other emission lines (see 
above).  It suggests that rotational modulation dominates the observed 
varibility, reaching about $\pm20-25$\% for \hbe\ and $\pm10$\% for \hal.  
The variations are however not as smooth as those of the accretion 
proxies described above (e.g., between phase 0.62 and 0.88 in Dec06), 
suggesting tha Balmer lines are subject to a higher level of instrinsic 
variability.  

Both \hal\ and \hbe\ exhibit Stokes $V$ signatures (not shown here) similar in 
shape to those of V2129~Oph (D07), with significant departures from antisymmetry.  
The corresponding longitudinal fields (listed in Table~\ref{tab:beff}), 
are peaking at about 0.23 and 1.4~kG for \hal\ and \hbe\ respectively;  they 
are smaller than, and correlate reasonably well with, those of other emission 
lines, suggesting that rotational modulation dominates the observed variations.  
The weaker longitudinal fields indicate that the circularly polarised signal from 
the accretion spots is more diluted (with the unpolarised light from the 
chromosphere and wind, and with weakly polarised contributions from the lower-field 
outer regions of accretion funnels) in Balmer lines than in the other accretion proxies;  
we speculate that most of the intrinsic variability observed in \hal\ and \hbe\ is due 
to this unpolarised chromospheric/wind component or to that from the outer regions of 
accretion funnels.   

We also note that 
the unpolarised profile of H$\alpha$ and H$\beta$ are much broader (with full 
widths at half maximum of about 240~\kms) than their Zeeman signatures (70~\kms\ wide).  
Moreover, both unpolarised lines are significantly blueshifted with respect to 
the stellar velocity rest frame (by as much as 15~\kms) while their Stokes $V$ 
signatures are slightly redshifted (by a few \kms, as \caii\ and \feii\ 
emission lines).  It confirms that Balmer lines contain several emission 
components, with at least a slightly redshifted one forming in accretion 
spots and producing the observed Stokes $V$ signatures, and a second (dominant) 
blueshifted one due to a chromosphere and/or a wind.  
This is all very reminiscent of what is observed on V2129~Oph (D07).  

\begin{figure}
\center{\includegraphics[scale=0.45,angle=-90]{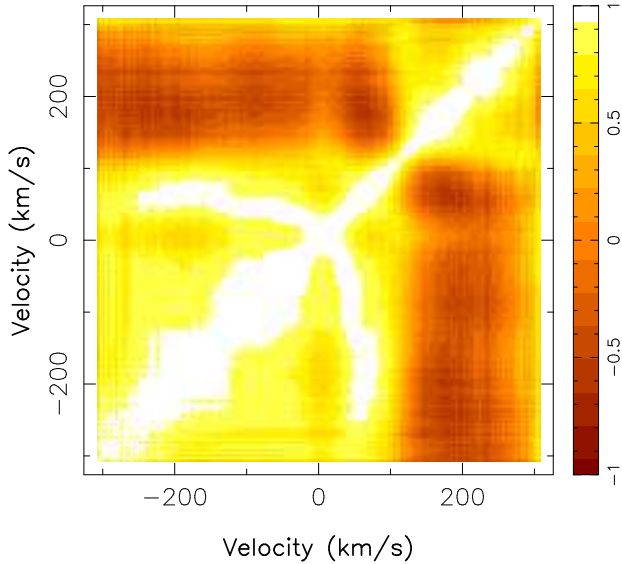}} 
\caption[]{H$\beta$ autocorrelation matrix computed from the Dec06 data.  White 
and dark brown indicate perfect correlation and anticorrelation respectively.  }
\label{fig:ccf}
\end{figure}

Balmer lines collected in Dec06 also include a conspicuous high-velocity 
component in their red wings (between $+100$ and $+300$~\kms, and mostly visible 
in H$\beta$).  This component varies roughly in phase opposition with the rest of 
the line and shows up as a hump around phase 1.63 (i.e., when the overall line 
emission is small) and as a relative dip (with respect to the average profile) 
around phase 1.26--1.38 (i.e., when line emission is large).  
This is readily visible on the corresponding H$\beta$ autocorrelation 
matrix (see Fig.~\ref{fig:ccf}).  By analogy with V2129~Oph (D07), we attribute 
it to disc material free falling onto the star within accretion funnels.  
Given the limited agreement between profiles taken almost one complete rotation 
cycle apart (e.g., phase 0.405 and 1.382, or phases 0.472 and 1.509), we 
conclude that component is apparently subject to a high level of intrinsic 
variability.  
This component is also present, though much fainter, in the Feb06 data set;  
it correlates well with the red wing of the broad \caii\ IRT emission 
component discussed above, thus suggesting a common origin.

\section{The accretion spots and magnetic topology}
\label{sec:zdi}

Now that the rotational modulation of Stokes $I$ and $V$ profiles of photospheric 
and emission lines is firmly established, we aim at modelling the observed 
modulation and derive from it maps of the accretion spots and the magnetic field at the 
surface of BP~Tau.  As in D07, we only use the LSD photospheric and \caii\ line 
profiles in this study.  Modelling \hei\ lines (and in particular the non-antisymmetric 
Stokes $V$ profiles, see Fig.~\ref{fig:he}) is rather uncertain;  it indeed requires 
additional independent information on the velocity fields and gradients within the 
line formation regions that we usually do not have and cannot easily access.  

\subsection{Model description}

The model we use here is directly inspired from that of D07.  It assumes that 
Stokes $V$ photospheric and \caii\ emission lines form in different regions of the 
stellar surface, reflecting the fact that accretion spots (where emission lines mostly 
form) coincide with dark cool spots at photospheric level (to which photospheric lines 
are mostly insensitive).  This model was fairly successful at reproducing observations 
of V2129~Oph.  Given the overall similarity of the present data sets with that of V2129~Oph 
(see above), we speculate that it should be adequate for BP~Tau as well;  in particular, 
it is compatible with reports that BP~Tau looks darkest in optical photometry 
when accretion spots are most visible \citep{Valenti03}.  

We assume that accretion lines can be described with the two component model introduced 
in Sec.~\ref{sec:rot}, combining a chromospheric emission component evenly
distributed over the star with an additional emission component concentrated
in local accretion  spots.   The emission-line Zeeman signatures are assumed to
be associated with the accretion component only.  In this model, rotational modulation 
of emission lines results from the accretion regions being carried in and out of the 
observer's view as the star rotates, the excess emission received from these regions 
being proportional to their projected area (as observations suggest).  

We introduce a vector magnetic field ${\bf B}$ and a local accretion
filling-factor $f$, describing the local sensitivity to accretion 
and photospheric lines\footnote{Note that the local accretion filling-factor we
define here is different from the usual accretion filling factor of the cTTS 
literature, i.e., the relative area of the total stellar surface covered by
accretion  spots.}.  For $f=0$, the local area on the protostar's surface
contributes fully to photospheric lines and generates no \caii\ excess
emission (and only unpolarised chromospheric \caii\ emission);  for $f=1$, the
local area does not contribute at all to photospheric lines and produces the
maximum amount of excess \caii\ emission and polarisation (in addition to the
unpolarised chromospheric \caii\ emission).   Spectral contributions for
intermediate values of $f$ are derived through linear combinations between the
$f=0$ and $f=1$ cases.  

We introduce one difference with the model of D07.  We assume that, in each 
local cell of the stellar surface, only a fraction $\psi$ of the cell area 
contains magnetic fields (whose local strength thus equals $B/\psi$).  We 
further assume that, within each surface cell, the accretion filling factor 
defined above only applies to the magnetic portion of the cell, while the 
non-magnetic portion corresponds to $f=0$.  We finally assume (for simplicity) that 
$\psi$ is constant over the whole stellar surface.  In this context, the model 
of D07 corresponds to the specific case where $\psi=1$.  Introducing $\psi$ 
brings the possibility of fitting magnetic fluxes ($B$) and magnetic 
strengths ($B/\psi$) independently, both having different effects in spectral lines 
of stars with low to moderate \vsini.  While the magnetic strength is mainly derived 
from fitting the far wings of the Zeeman signatures, the magnetic flux is obtained 
essentially by adjusting the amplitude of the Stokes $V$ profiles\footnote{Magnetic 
distortions on Stokes $I$ profiles remain small, at least on the average photospheric 
line considered here, much smaller in particular than the observed rotational 
modulation described in Sec.~\ref{sec:rot}.}.  
We consider $\psi$, called the magnetic filling factor, as a free parameter 
that we optimise by minimising the amount of information in the magnetic 
image (see below).  

In this context, the local synthetic photospheric and emission Stokes $I$ and 
$V$ lines profiles emerging from each grid cell (noted respectively \Ip, 
\Vp, \Ie\ and \Ve) are given by:  
\begin{eqnarray}
\Ip & = &  (1-f) \psi \Im + (1-\psi) \Iq  \\ 
\Vp & = &  (1-f) \psi \Vm \\ 
\Ie & = & \Ic  + f \psi (\Ia -1)  \\ 
\Ve & = & f \psi \Va 
\label{eq:mod}
\end{eqnarray}
where \Im\ and \Iq\ are the local Stokes $I$ photospheric profiles  
corresponding to the magnetic and non-magnetic areas, \Vm\ the local 
Stokes $V$ photospheric profile corresponding to the magnetic areas, 
\Ic\ and \Ia\ the local Stokes $I$ profiles corresponding to the chromospheric 
and accretion emission components, and \Va\ the local Stokes $V$ profile 
corresponding to the accretion emission component.  

We obtain both $B$ and $f$ by fitting the corresponding
synthetic Stokes $V$ and $I$ spectra to the observed Zeeman signatures and 
unpolarised LSD profiles of photospheric lines and \caii\ emission cores.  
The code we use for fitting $B$ and $f$ is adapted from the stellar surface
magnetic imaging code of \citet{Donati01b} and \citet{Donati06b}.  The field 
is decomposed in its poloidal and toroidal components (with the poloidal field 
split itself between its radial and non radial contributions) and described 
as a spherical harmonics expansion, whose coefficients ($\alpha_{\ell,m}$,  
$\beta_{\ell,m}$ and $\gamma_{\ell,m}$ for the radial, non-radial poloidal 
and toroidal components respectively, where $\ell$ and $m$ denote the modes 
orders and degrees, see \citealt{Donati06b} for more details) are unknowns 
in the fitting procedure.  The remaining free parameters are the values $f_j$ 
of $f$ over the surface of the star (divided into a grid of thousands of small 
surface pixels indexed with $j$).  

The inversion problem being ill-posed, we use the principles of maximum entropy 
image reconstruction to make the solution unique \citep{Skilling84}.  The 
entropy function, computed from $f_j$, $\alpha_{\ell,m}$, $\beta_{\ell,m}$ and 
$\gamma_{\ell,m}$, allows us to select the image with minimum information (maximum 
entropy) given a predetermined quality of the fit to the data, usually set to a 
reduced chi-square \chisqr\ of about 1.  By fitting only $\alpha_{\ell,m}$ and 
$\beta_{\ell,m}$, we have the possibility of trying to force the solution 
towards a purely poloidal field;  similarly, by weighting odd or even 
coefficients very heavily in their contribution to the entropy function, we 
also have the option of driving the recovered field topology towards 
symmetry or antisymmetry with respect to the centre of the star.  Another 
significant advantage of this new imaging method is that it suffers much smaller 
crosstalk between field components than the original one \citep{Donati97b, Donati01b}.  

We use Unno-Rachkovsky's equations \citep[e.g.,][]{Landi04} to model the LSD profiles 
of local photospheric lines (i.e., \Iq, \Im\ and \Vm), whose equivalent wavelength and 
Land\'e factors are set to 
620~nm and 1.2 respectively.  We adjust the free Milne-Eddington model parameters by 
fitting Unno-Rachkovsky's equations to an unpolarised LSD profile of the slowly 
rotating standard star $\delta$~Eri whose spectral type is very similar to that of 
BP~Tau.  We then derive both the rotational broadening \vsini\ and the radial 
velocity \vrad\ of BP~Tau by fitting our series of LSD Stokes $I$ profiles and 
selecting the values that minimise the image information content (for a given 
quality of the fit to the data).  We find $\vsini=9.0\pm0.5$~\kms\ and 
$\vrad=15.5\pm0.5$~\kms, in good agreement with previous estimates 
\citep[e.g.,][]{Johns99b}.  

The local \caii\ Stokes $I$ chromospheric and accretion profiles \Ic\ and \Ia\ are 
both described (in the absence of magnetic fields) with a simple gaussian centred at 
850~nm, whose full width at half maximum (about 15~\kms) is derived from a fit to 
the observed \caii\ emission profiles of BP~Tau (assuming the rotational broadening 
obtained in the previous modelling step).  The effect of magnetic fields on \Ia\ 
and the corresponding Stokes $I$ and $V$ profiles are 
modelled using Unno-Rachkovsky's equations and assuming a unit Land\'e factor.  
The equivalent width ratio of the accretion emission component \Ia\ to that of the 
quiet chromospheric component \Ic\ is set to 4/$\psi$ (the division by $\psi$ ensuring 
that the total amount of local emission, and hence the derived accretion filling 
factor map, does not depend on $\psi$);  similar results being obtained for other 
values, e.g., 3/$\psi$ or 5/$\psi$.  The equivalent width $\epsilon_{\rm b}$ 
of the quiet chromospheric emission component \Ic\ is a free parameter 
that we optimise by minimising 
the information content of the magnetic image (see below);  increasing $\epsilon_{\rm b}$ 
will force the imaging code to decrease the relative fraction of accretion emission with 
respect to that of chromospheric emission (and vice versa), so that the resulting 
synthetic emission profiles match the observed ones.

\begin{figure}
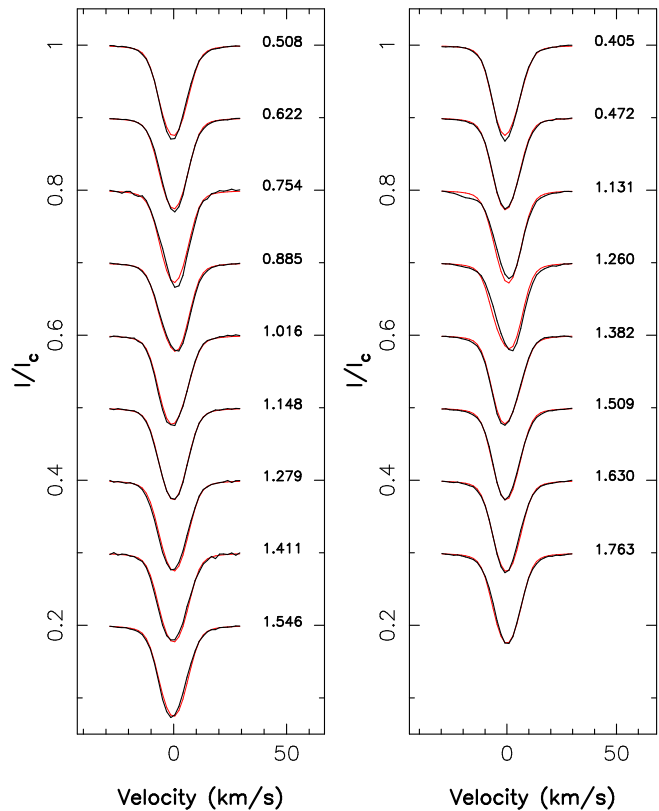

\center{\hbox{\includegraphics[scale=0.58,angle=-90]{fig/bptau_fiti1.ps}\hspace{1mm}
              \includegraphics[scale=0.58,angle=-90]{fig/bptau_fiti2.ps}}}
\caption[]{Stokes $I$ LSD profiles of photospheric lines of BP~Tau (thick black 
line) along with the maximum entropy fit (thin red 
line) to the data, for both Feb06 (left) and Dec06 (right) runs.  
The rotational cycle of each observation is noted next to each profile.  }
\label{fig:fiti}
\end{figure}

\begin{figure}
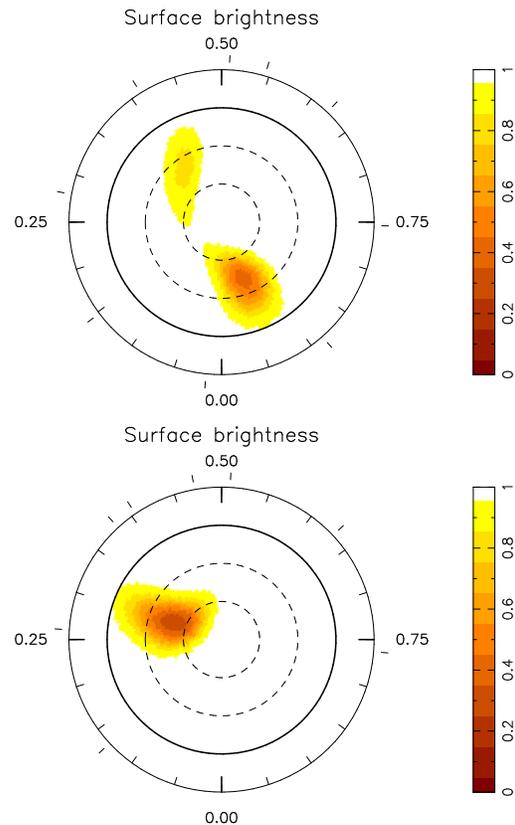

\center{\includegraphics[scale=0.40]{fig/bptau_mapi1.ps}}
\center{\includegraphics[scale=0.40]{fig/bptau_mapi2.ps}}
\caption[]{Maps of the local surface brightness (relative to that of the quiet
photosphere) on BP~Tau, for both Feb06 (top) and Dec06 (bottom) runs.  
The star is shown in flattened polar projection down
to latitudes of  $-30\degr$, with the equator depicted as a bold circle and
parallels as dashed circles.   Radial ticks around each plot indicate the
phases of observation.  } 
\label{fig:mapi}
\end{figure}

\subsection{First attempt}

The first attempt consists in fitting, for both data sets, LSD Stokes $I$ photospheric 
profiles only.  In addition to provide an accurate estimate of both \vsini\ and \vrad\ 
(see above), this step allows us to check the validity of our assumption that the shape of 
photospheric lines is mostly distorted by dark spots at the surface of the star and that 
the observed profile variability is compatible with rotational modulation.  It also 
enables us to check a second modelling assumption, i.e., that the putative dark 
photospheric spots at the surface of the star are indeed more or less coincident with 
the accretion regions identified from the intensity and radial velocity modulation of 
emission lines (see Sec.~\ref{sec:rot}).  

We start by removing veiling from all LSD Stokes $I$ photospheric profiles, i.e., by scaling 
them to the same equivalent width;  modelling veiling as part of the imaging process appears 
undesirable at this stage given that veiling only weakly correlates with the other 
parameters we aim at modelling (and in particular the Zeeman signatures and the shape 
and strength of photospheric and emission line profiles).   
The fit we obtain (see Fig.~\ref{fig:fiti}) matches the data at a \sn\ level 
of about 500 and yields a $\chisqr$ improvement of about a factor of 2 with 
respect to synthetic profiles corresponding to an unspotted star.  Most of the main 
profile distortions, like for instance the profile asymmetries at phase 1.260 and 
1.382 in the Dec06 data set (see Sec.~\ref{sec:rot} and Fig.~\ref{fig:pho}), are 
reproduced by our model, leading us to conclude that our series of Stokes $I$ 
profiles of BP~Tau are compatible with rotational modulation induced by dark 
surface spots.  The reconstructed images (see Fig.~\ref{fig:mapi}) show one main spot 
at each epoch, covering in both cases about 2\% of the total stellar surface.  

These images should not be viewed as true brightness images of the photosphere of 
BP~Tau;  given the rather low \vsini\ (by Doppler imaging standards) and the subsequently 
limited spatial resolution (about 0.1 cycle at the rotation equator), only the largest 
non-axisymmetric spots are reconstructed here.  In particular, polar spots such as that 
detected on V2129~Oph (producing no rotational modulation) are hardy detectable on 
BP~Tau through Stokes $I$ profiles only.  The spots we recover should thus be 
seen as the most reliable features without which the distortions and variability of 
unpolarised profiles cannot be interpreted in terms of rotational modulation.  

We find that the reconstructed dark spots are reasonably close to where we expect 
accretion regions to concentrate, i.e., at high latitudes -- given the low amplitude 
of velocity variations of emission lines (see Fig.~\ref{fig:ewblonvrad}) -- and at phase 1.02 
(in Feb06) and 1.38 (in Dec06) -- at which emission lines show maximum emission and 
reach a median position in their blue to red transit (see Figs.~\ref{fig:irt}, 
\ref{fig:he} and \ref{fig:ewblonvrad}).  They apparently lag slightly behind (by about 0.05 
to 0.10 rotation cycle) the accretion spots themselves.  Although the spatial resolution 
is likely too low to ascertain the latter point, it is nevertheless enough to 
confirm our basic working assumptions.  

At this stage, one may argue that that the rotational broadening we derive in this 
process is overestimated as it neglects line broadening induced by potentially strong 
magnetic fields at the surface of BP~Tau;  moreover, some of the profile distortions 
we trace and attribute to dark spots could be due to magnetic fields instead.  We 
however think that is not the case, for at least two main reasons.  Firstly, the 
\vsini\ we derive is fully compatible (and even slightly lower) than that of 
\citet{Johns99b}, demonstrating that our estimate is obviously not an upper limit.  
Secondly, we would expect to see significant changes with rotational phase in the 
width of the LSD unpolarised profile (as we do for instance in chemically peculiar 
stars hosting strong magnetic fields, \citealt[e.g.,][]{Landstreet88}) since the 
magnetic topology of BP~Tau is apparently not fully axisymmetric;  no such effect 
is detected.  

Yet, the magnetic fields detected on BP~Tau, reaching strengths of up to 8~kG (see 
Table~\ref{tab:beff}), are in principle strong enough to produce straightforwardly 
visible broadening of unpolarised lines.  So why do we see no such signatures in our 
series of Stokes $I$ LSD profiles?  This is because very intense fields actually 
concentrate where accretion is strongest and thus hide in very dark photospheric 
regions;  for this reason, they mainly show up in emission lines and almost disappear 
from optical photospheric lines, whose widths therefore remain mostly unaffected.  
This interpretation (mostly speculative at this point) is confirmed below.  

\begin{figure}
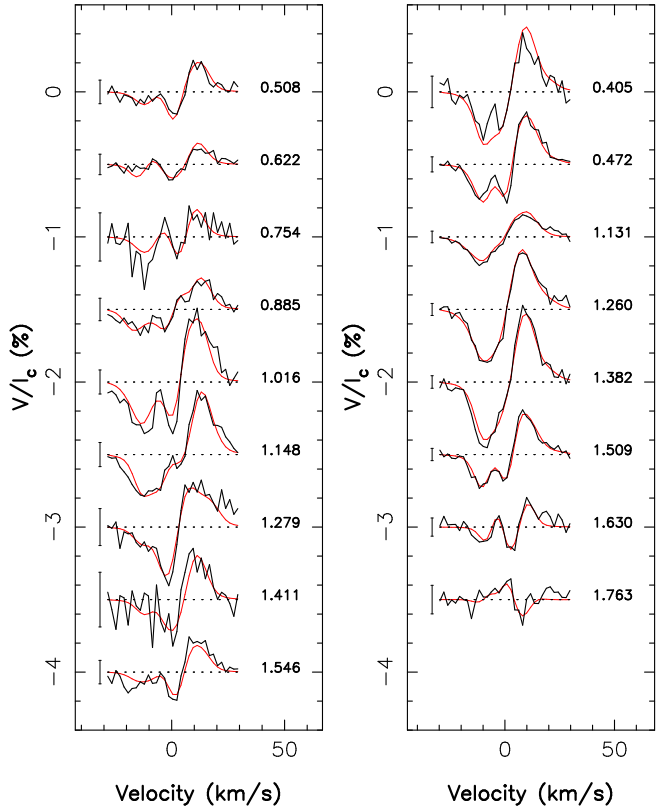

\center{\hbox{\includegraphics[scale=0.58,angle=-90]{fig/bptau_fitv1.ps}\hspace{1mm}
              \includegraphics[scale=0.58,angle=-90]{fig/bptau_fitv2.ps}}}
\caption[]{Stokes $V$ LSD profiles of photospheric lines of BP~Tau 
(thick black line) along with the maximum entropy fit (thin red line) to the data, for 
both Feb06 (left) and Dec06 (right) runs.  
The rotational cycle of each observation and 3$\sigma$ error bars are also shown 
next to each profile.  } 
\label{fig:fitv}
\end{figure}

\begin{figure}
\center{\hbox{\includegraphics[scale=0.58,angle=-90]{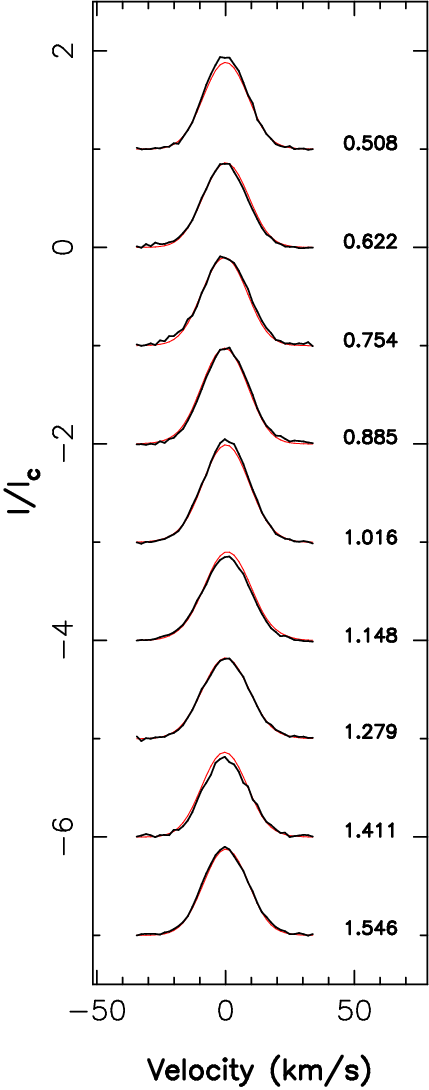}\hspace{1mm}
              \includegraphics[scale=0.58,angle=-90]{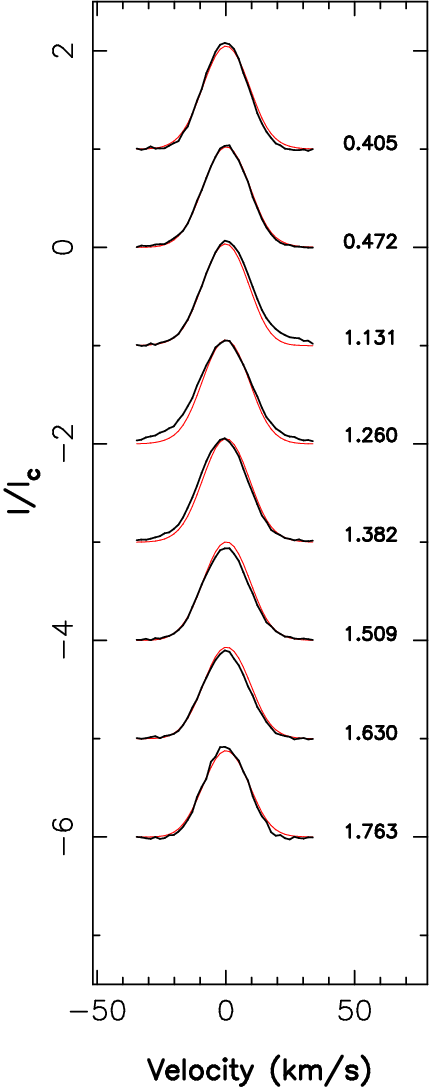}}}
\caption[]{The same as Fig.~\ref{fig:fiti} but for the \caii\ emission profiles of 
BP~Tau. } 
\label{fig:fitci}
\end{figure}

\begin{figure}
\center{\hbox{\includegraphics[scale=0.58,angle=-90]{fig/bptau_fitcv1.ps}\hspace{1mm}
              \includegraphics[scale=0.58,angle=-90]{fig/bptau_fitcv2.ps}}}
\caption[]{The same as Fig.~\ref{fig:fitv} but for the \caii\ emission profiles of 
BP~Tau. } 
\label{fig:fitcv}
\end{figure}

\begin{figure}
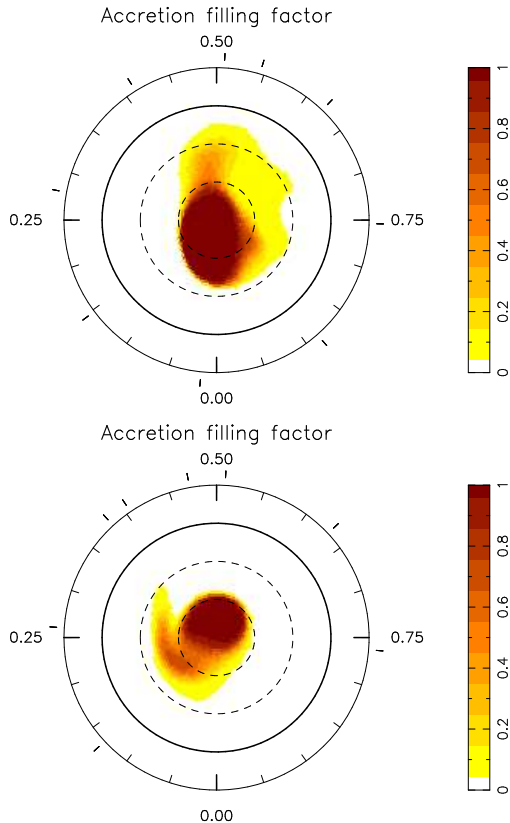

\center{\includegraphics[scale=0.40]{fig/bptau_mapf1.ps}} 
\center{\includegraphics[scale=0.40]{fig/bptau_mapf2.ps}} 
\caption[]{Maps of the local accretion filling factor $f$ on BP~Tau, for both Feb06 (top) 
and Dec06 (bottom) runs.   Note that these maps also depict the reconstructed 
photospheric brightness (equal to  $1-\psi f$), our model being based on the 
assumption that dark photospheric spots and accretion regions coincide.  These maps, 
obtained when modelling all data simultaneously, can be compared to those of Fig.~\ref{fig:mapi} 
derived from fitting Stokes $I$ profiles of photospheric lines only.   }  
\label{fig:mapf}
\end{figure}

\begin{figure*}
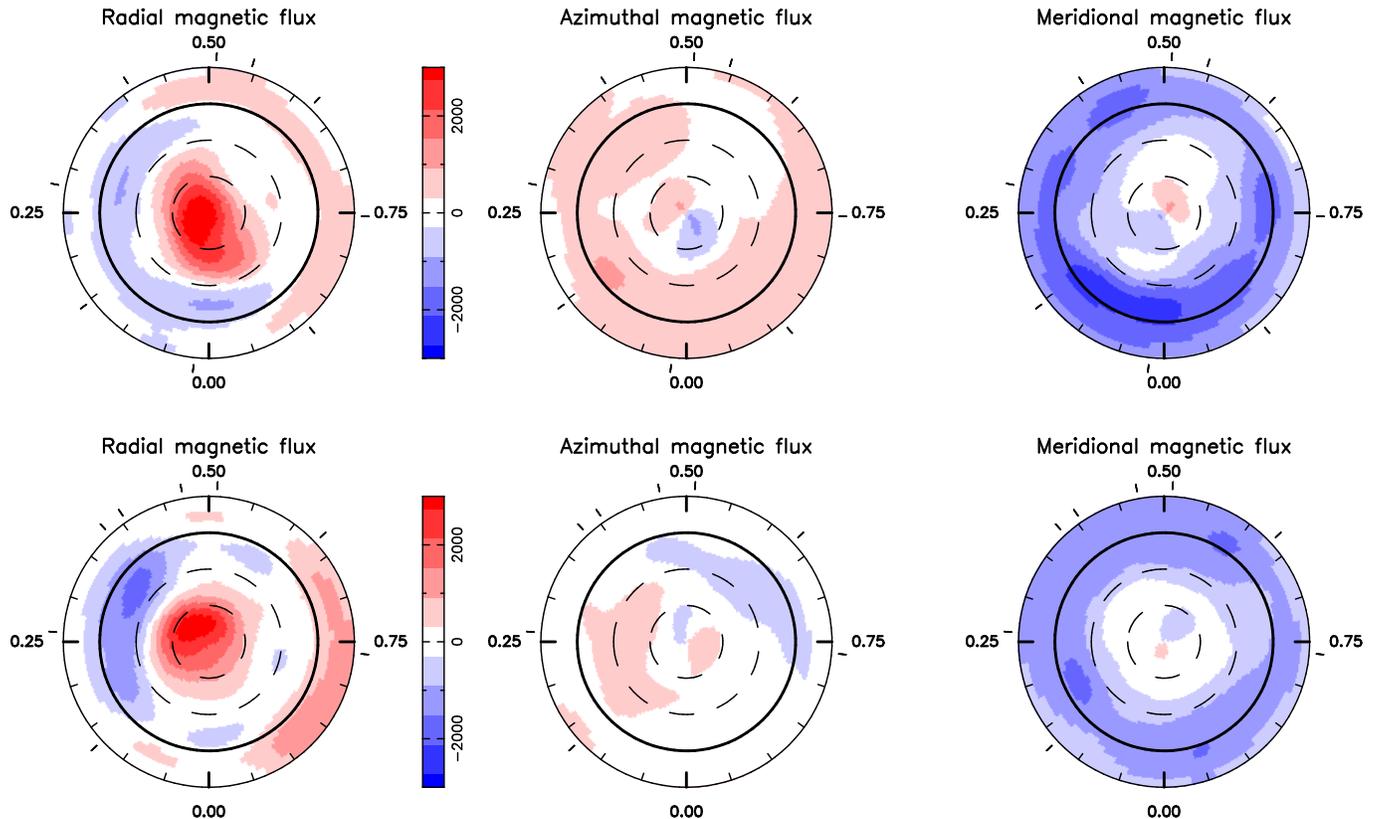

\center{\includegraphics[scale=0.75]{fig/bptau_mapb1_new.ps}}
\center{\includegraphics[scale=0.75]{fig/bptau_mapb2_new.ps}}
\caption[]{Magnetic topologies of BP~Tau in Feb06 (top) and Dec06 (bottom), 
reconstructed from a simultaneous fit to the complete series of Stokes $I$ and 
$V$ profiles of photospheric lines and \caii\ emission cores.  The three components 
of the field in spherical coordinates are displayed (from left to right), with 
magnetic fluxes labelled in G.  } 
\label{fig:mapv}
\end{figure*}

\subsection{Complete modelling}

In this second step, we carry out the complete modelling by fitting, for each observing 
epoch, a magnetic topology and an accretion filling-factor map to all Stokes $I$ and $V$ 
profiles of photospheric and emission lines simultaneously.  The remaining free parameters 
to adjust in this process are the magnetic filling factor $\psi$ (arbitrarily set to 1 in 
the first modelling step) and the equivalent width $\epsilon_{\rm b}$ of the quiet 
chromospheric emission component.  We 
find that $\psi$ needs to be significantly smaller than 1 for LSD Stokes $V$ profiles of 
photospheric lines to be fitted down to noise level, in particular in the far line wings.  

Given the available spatial resolution in the data, spherical harmonics expansions 
describing the magnetic field were truncated at $\ell=10$.  Moreover, the field 
reconstruction was oriented towards antisymmetric (rather than symmetric) magnetic 
topologies (about the center of star), by favouring spherical harmonics terms with odd 
$\ell$ values (see D07);  dominantly antisymmetric field configurations are indeed the only 
ones capable of yielding (through their dipolar component in particular) the exclusive 
high-latitude anchoring of accretion funnels that we observe for BP~Tau 
\citep[e.g.,][]{Gregory06}.  

The value that minimises the amount of energy in the reconstructed field given a unit 
\chisqr\ fit to the data is $\psi=0.25$.  We also find that $\epsilon_{\rm b}$ should be 
about 60\% of the average observed \caii\ line emission, the remaining 
40\% being produced by accretion spots and modulated by rotation.  
The final fits to the data, corresponding to a unit \chisqr,  are shown in 
Fig.~\ref{fig:fitv} for the Stokes $V$ LSD photospheric profiles, in Fig.~\ref{fig:fitci} 
for the Stokes $I$ \caii\ emission lines and in Fig.~\ref{fig:fitcv} for the Stokes $V$ 
\caii\ emission lines;  the new fit to the Stokes $I$ LSD photospheric profiles (not 
shown) is only slightly worse than that of Fig.~\ref{fig:fiti}.  The reconstructed 
accretion filling factor and magnetic flux maps are 
shown in Figs.~\ref{fig:mapf} and 
\ref{fig:mapv} respectively.  Since accretions regions are assumed to concide with dark 
photospheric spots in our model, Fig~\ref{fig:mapf} also provides a description of the 
reconstructed photospheric brightness.  

The updated \vsini\ estimate we derive (as a by-product) from this new modelling is virtually 
identical to that obtained in the previous modelling step (involving no magnetic fields);  
it confirms in particular that magnetic distortions on Stokes $I$ profiles are weak (at 
least on the average line that we consider here).  
Attempts at fitting the data (and in particular the Stokes 
$V$ sets) assuming different values of the rotation period confirm that optimal results 
are obtained for a rotation period close to the nominal value of 7.6~d (within about 0.5~d).  

The reconstructed accretion regions (and dark photospheric spots) are mostly located over 
the polar regions of BP~Tau;  
each spread out on about 8\% of the star, actually covering only 2\% of the total surface 
given than only one fourth of each surface pixel is subject to accretion and hosts magnetic 
fields ($\psi=0.25$, see model description above).  
While not identical to the initial brightness maps of Fig.~\ref{fig:mapi} (derived from 
fitting Stokes $I$ profiles of photospheric lines alone in our first modelling attempt), 
the new accretion filling factor (and photospheric brightness) maps that we now obtain 
from fitting the full data set are nevertheless similar;  in particular both feature mid-latitude 
appendages located at more or less the same phase as the brightness spots of 
Fig.~\ref{fig:mapi}.  We suspect that some of the discrepancy between both sets of maps 
is due to the fact that the assumptions underlying our model (e.g., dark spots overlapping 
accretion regions) are too simple;  however, since they both yield acceptable fits to the 
Stokes $I$ LSD profiles, we conclude that their differences likely reflect the limited spatial 
resolution available in the spectra of BP~Tau.  Our result confirms that the magnetic 
broadening of Stokes $I$ LSD profiles is small, thereby confirming the preliminary 
conclusions reached above.  

The magnetic topologies we recover at both epochs are similar, apart from a phase shift 
of about 0.25 rotation cycle.  They both include one main positive radial field feature 
close to the magnetic pole, where the magnetic flux reaches 3~kG and the field strength 
up to 12~kG (since $\psi=0.25$).  The average field strength over the accretion regions 
is about 9~kG, in agreement with the highest values of longitudinal fields traced by 
emission lines.  In non accreting regions, the average magnetic flux is about 1.2~kG;  
the field is dominantly meridional and pointing away from the observer, matching the 
constantly negative longitudinal fields traced by photospheric lines.  

We find that the magnetic topologies we recover are dominantly poloidal, with only about 10\% 
of the magnetic energy concentrating in the toroidal component.  In particular, the toroidal 
field of BP~Tau is definitely smaller (in terms of fractional magnetic energy) than that of 
V2129~Oph, which gathered about 20\% of the reconstructed magnetic energy (D07).  This toroidal 
component seems nonetheless real;  while fitting the Stokes $V$ data at unit \chisqr\ with a 
purely poloidal field is possible, the magnetic field we recover is significantly stronger 
suggesting that a purely poloidal solution is far less likely (according to maximum entropy 
principles).  

The spherical harmonics terms dominating the recovered magnetic topology corresponds to a 
slightly tilted dipole, with about 50\% of the magnetic energy concentrating in 
$\ell=1$ modes;  the corresponding dipole strength is about 1.2~kG.  The second dominant 
term is a slightly tilted octupole, with about 30\% of the magnetic energy gathering in 
$\ell=3$ modes;  the corresponding octupole strength is about 1.6~kG\footnote{While the 
dipole and the octupole are both tilted by about 10\degr\ with respect to the rotation axis, 
they are not tilted towards the 
same phase so that their poles do not coincide.}.  This makes BP~Tau 
fairly different from V2129~Oph, whose field comprises a dominantly octupolar large-scale 
magnetic field and only a small dipole component (D07).  All remaining spherical harmonics 
terms of the large-scale (i.e., $\ell<10$) poloidal field of BP~Tau contain altogether no 
more than 10\% of the total magnetic energy.  Note that our data does not exclude the potential 
presence of small-scale magnetic features (producing no detectable polarisation signatures) 
at the surface of BP~Tau.  

Although small, the intrinsic differences between the two reconstructed magnetic topologies 
of BP~Tau are apparently real and can be traced directly to differences in the data sets 
themselves;  Zeeman signatures in photospheric lines around phases of maximum longitudinal 
fields (i.e., at cycles 1.016 and 1.148 in Feb06 and 1.260 and 1.382 in Dec06) are indeed 
different in shape and undergo different temporal evolution.  These changes are however 
fairly limited and do not affect the gross characteristics of the large-scale field.

\section{Magnetospheric accretion and corona}
\label{sec:mag}

As for V2129~Oph, we propose here an illustration of how accretion may proceed between 
the inner disc and the surface of the  star.  To do this, we extrapolate the reconstructed 
magnetic field over the whole magnetosphere, assuming the 3D field topology is mainly 
potential and becomes radial beyond a certain magnetospheric radius \rmag\ from
the star (to mimic the opening of the largest magnetic loops under the coronal
pressure, \citealt{Jardine02, Jardine06, Gregory06}).   In non-accreting stars,
this distance is usually assumed to be smaller than or equal to the corotation
radius (\rcor) at which the Keplerian orbital period equals the stellar
rotation period.  In cTTSs however, the magnetic field of the  protostar is
presumably clearing out the central part of the accretion disc and extending 
as far as the inner disc rim.   

\begin{figure*}
\center{\includegraphics[scale=0.40]{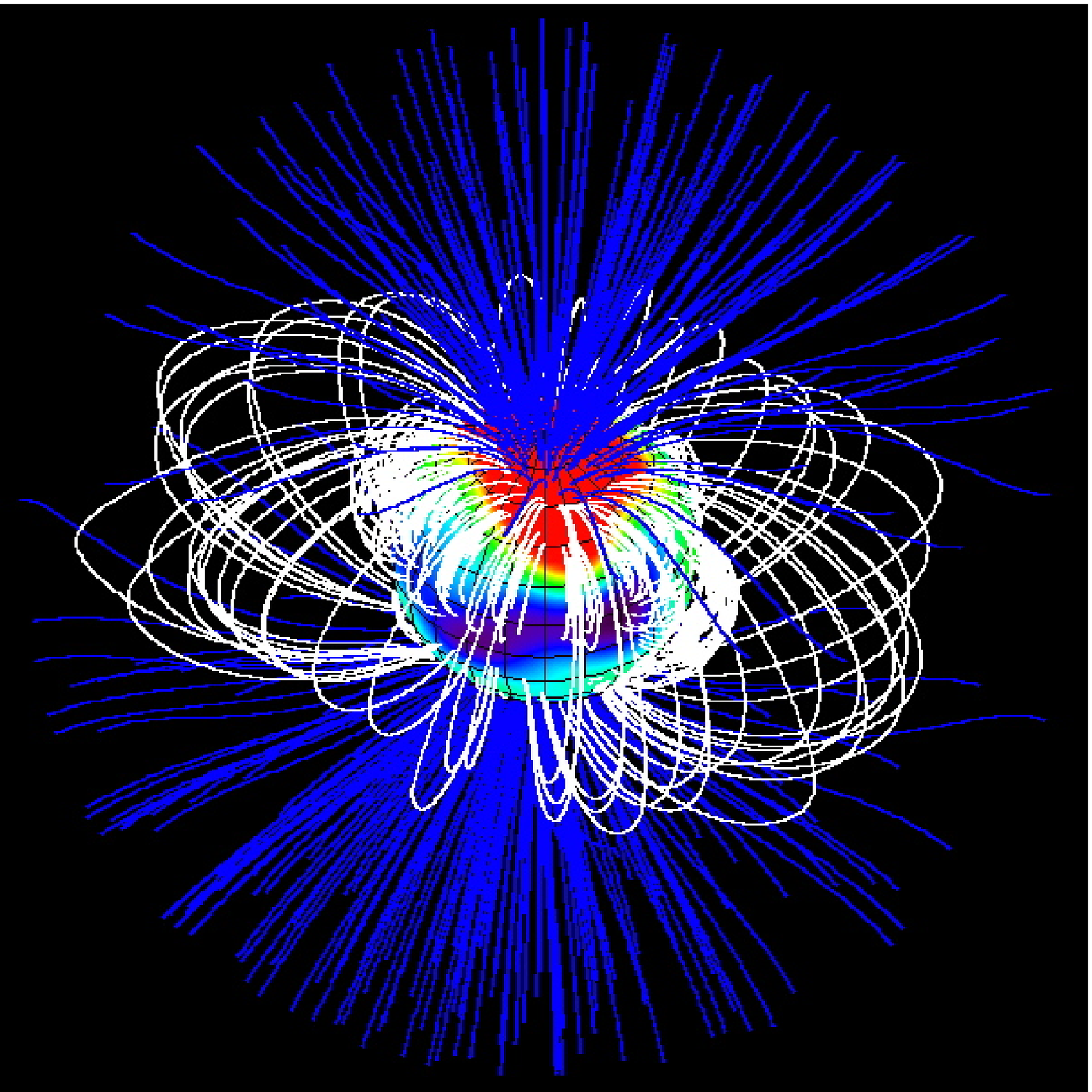}
        \includegraphics[scale=0.40]{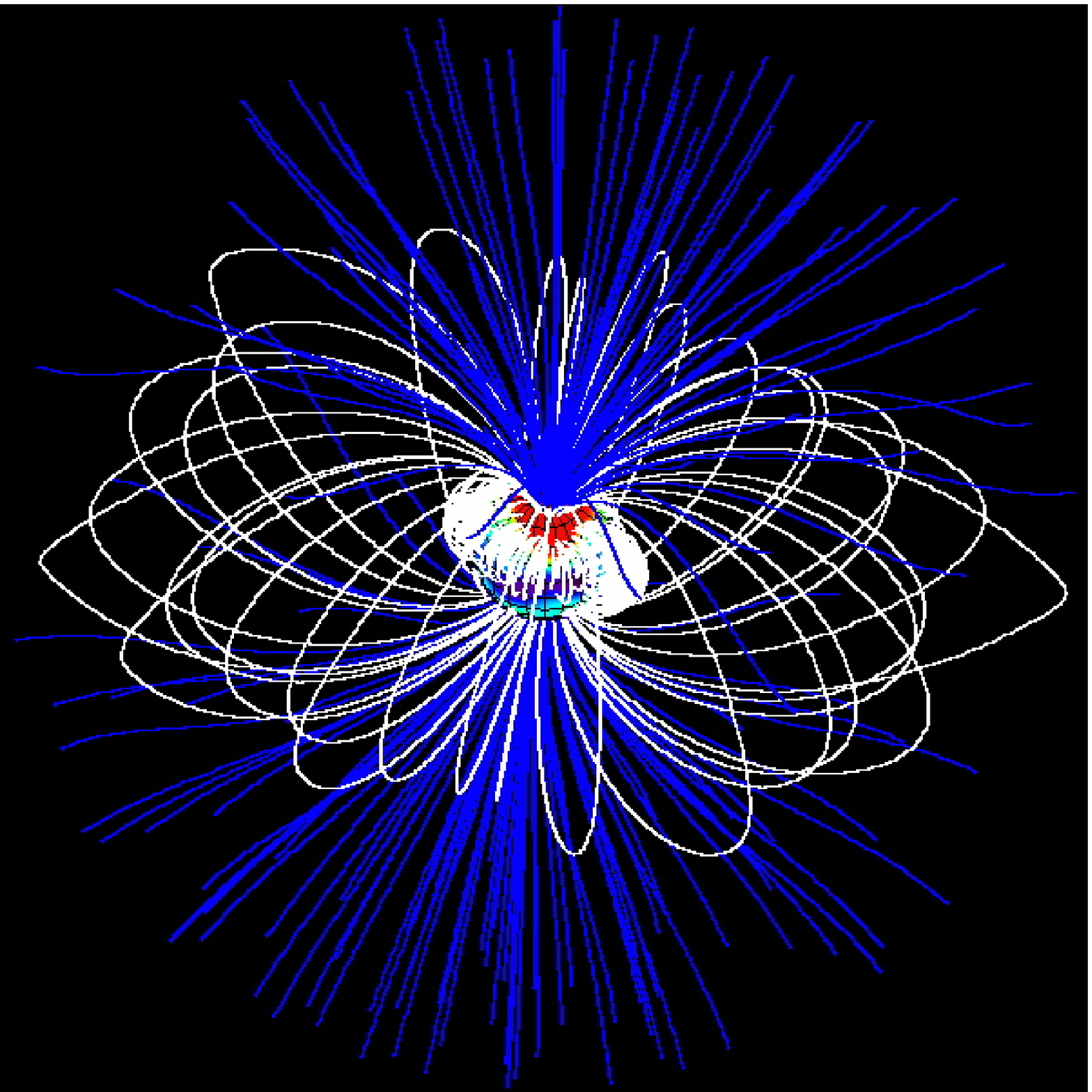}}
\caption[]{Magnetospheric topology of BP~Tau as derived from potential
extrapolations of the Feb06 surface magnetic field distribution (top panel 
of Fig.~\ref{fig:mapv}).   The magnetosphere is assumed to extend up to the inner
disc radius, equal to  3.5 and 7.5~\rstar\ in the left and right panels
respectively.  The complex  magnetic topology close to the surface of the star
is very obvious.   In both cases, the star is shown at rotational phase 0.0.
The colour patches  at the surface of the star represent the radial component
of the field  (with red and blue corresponding to positive and negative
polarities);  open  and closed field lines are shown in blue and white
respectively.  }
\label{fig:mag}
\end{figure*}

This is of course only an approximation.  In particular, the strong plasma flows
linking the disc to the stellar surface as a result of mass-accretion will likely 
prevent the field from being potential by building up a strong azimuthal component in the 
magnetosphere.  Magnetic tracers at the base of accretion funnels (and in particular 
\hei\ emission lines, see Sec.~\ref{sec:rot}) suggest that this is indeed the case.  
We nevertheless use this approach as a first step;  further detailed simulations are 
postponed to forthcoming papers.  The magnetospheric maps we derive for BP~Tau (see 
Fig.~\ref{fig:mag}
for two possible values of \rmag\ at epoch Feb06) show that the field topology is
complex close to the stellar surface, but dominated by more extended
open and closed field lines at larger distances.  

From these maps, we can estimate where the accretion funnels are located and where 
they are anchored at the surface of the star.  We do this by identifying those 
magnetospheric field lines that are able to accrete material from the disc, i.e., that 
link the star to the disc and intersect the rotational equator with effective
gravity pointing inwards in the co-rotating frame of reference 
\citep[e.g.][]{Gregory06, Gregory07}.  We find that accretion spots concentrate at 
high latitudes if we assume $\rmag=7.5$~\rstar\ (see Fig.~\ref{fig:ftp}).  For values 
of \rmag\ smaller than about 3.5~\rstar, equatorial accretion spots start to form;  
most of the accretion occurs onto the equator when $\rmag\leq2.5$~\rstar.  

\begin{figure*}
\center{\hbox{\includegraphics[scale=0.48]{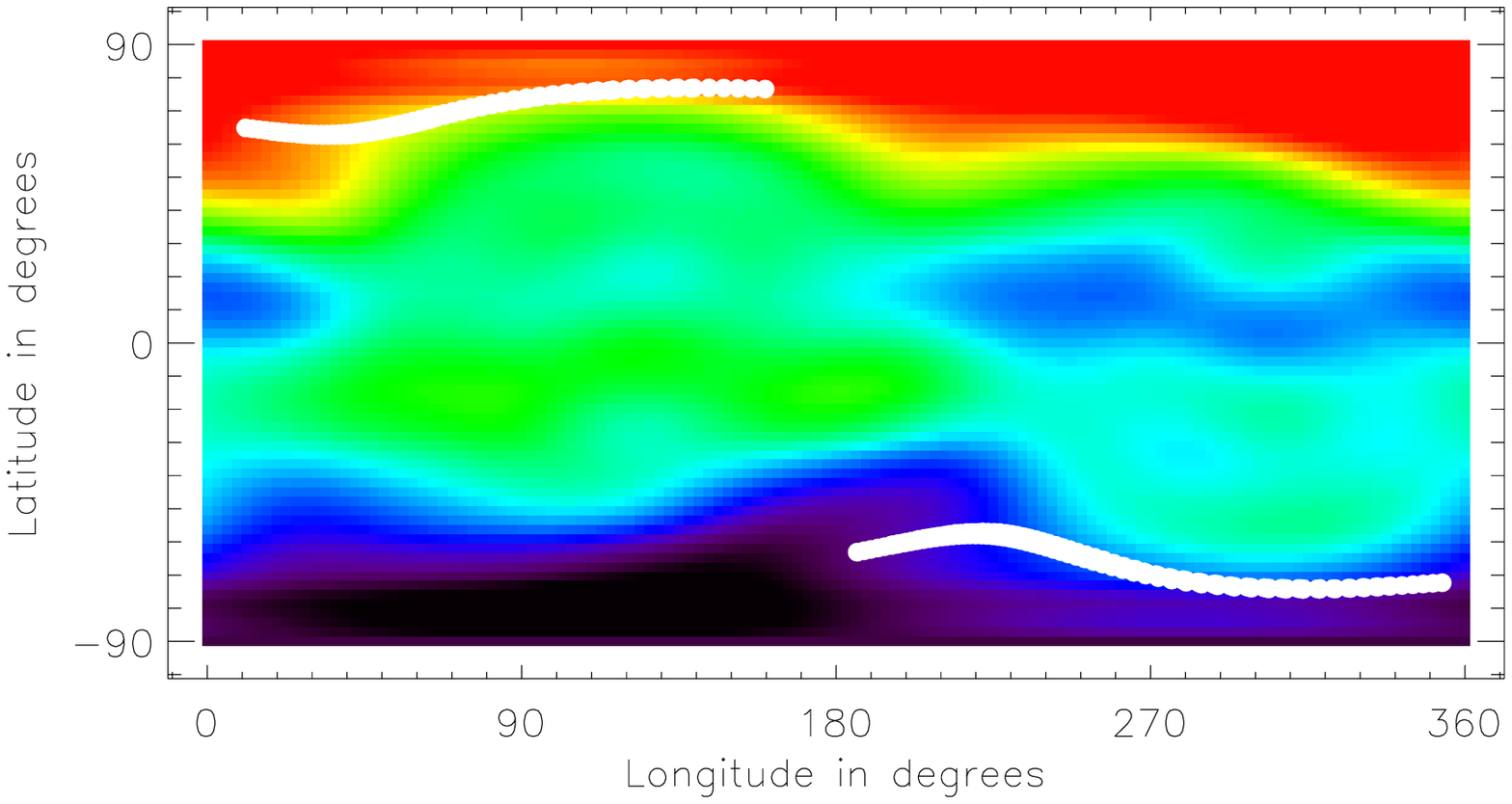}
              \includegraphics[scale=0.48]{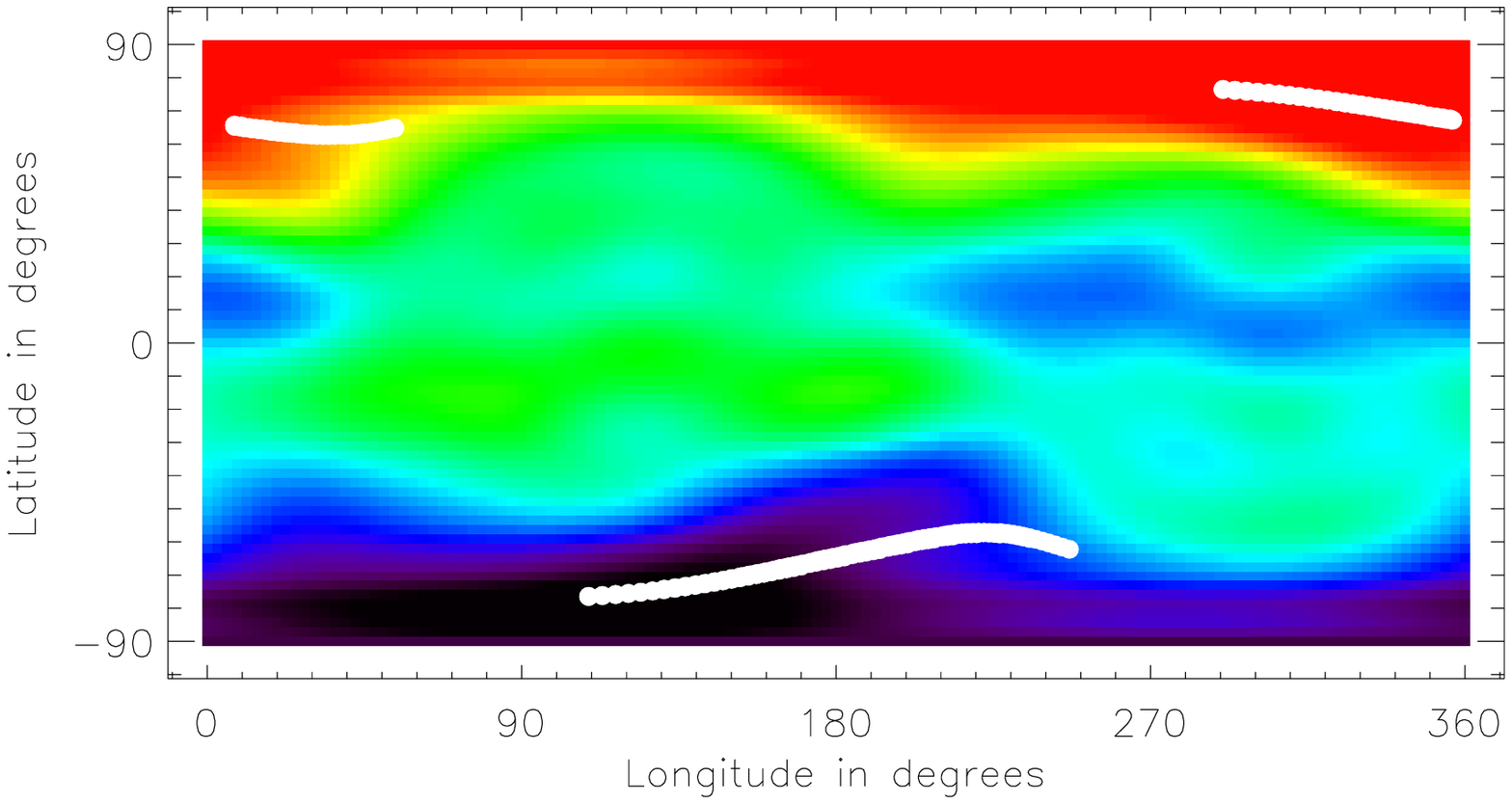}}}  
\caption[]{Location of the footpoints of accreting field lines at the surface
of the BP~Tau (white circles), assuming that the magnetosphere (extrapolated 
from the Feb06 magnetic maps) extends to 7.5~\rstar, and that the accretion disc 
coincides with the stellar rotation equator (no magnetic warp, left panel) or with 
the stellar dipole magnetic equator (magnetic warp, right panel).  
The colours at the surface of the star depict  the radial field component (with
red and blue corresponding to positive and negative  polarities).  Note that
phase runs backwards with longitude, i.e., decreases from 1 to 0  while
longitude increases from 0\degr\ to 360\degr.  }
\label{fig:ftp}
\end{figure*}

Given the results of Sec.~\ref{sec:zdi}, demonstrating that accretion spots are 
located at high latitude, we conclude that \rmag\ is at least equal to about 4~\rstar\  
and may extend as far out as \rcor, equal to about 0.07~AU or 7.5~\rstar\ for 
BP~Tau\footnote{Assuming that the sporadic fluctuations in the period of photometric 
variations \citep[e.g.,][]{Simon90} is due to contamination by optical light from the 
inner disc rim (presumably modulated on a timescale corresponding to the local 
Keplerian period), it would imply that the radius of the inner disc, equal to \rmag, 
also fluctuates with time, from about 6.5 to 8.5~\rstar\ typically.}.  
Our modelling also demonstrates that accretion spots concentrate at phase of maximum 
line emission (0.0 in Feb06 and 0.4 in Dec06, see Fig.~\ref{fig:mapf});  this is 
further confirmed by the transient absorption appearing in the red wing of Balmer 
lines at these phases (see Sec.~\ref{sec:rot}).  When assuming that the accretion disc lies 
within the equatorial rotation plane, the northern accretion hot spot that 
our extrapolated magnetosphere model predicts (located at longitude 90\degr\ or phase 
0.75 in Feb06, see Fig.~\ref{fig:ftp} left panel) does not match the 
observations.  Assuming that the accretion disc is magnetically warped around \rmag\ and 
locally lies within the magnetic equator of the large-scale dipole component produces a 
better agreement with observations (see Fig.~\ref{fig:ftp} right panel).  


\section{Discussion}
\label{sec:dis}

The new spectropolarimetric data we collected on BP~Tau at 2 different epochs enabled us 
to obtain a realistic model of the large-scale magnetic field on this prototypical cTTS.  
Comparing our results with those recently obtained for a more massive cTTS (V2129~Oph, D07) 
suggests new hints on how magnetic fields are produced in young low-mass stars and how 
they make young stars interact with their accretion discs.  

\subsection{Dark spots and accretion regions} 

As for V2129~Oph, we observe that photospheric lines and the narrow emission lines forming 
at the footpoints of accretion funnels (e.g.\ \caii\ IRT, \hei\ and \feii\ emission lines) 
are mainly modulated by rotation.  Broad emission lines (e.g., Balmer lines or the broad 
emission component of the IRT) are also modulated by rotation but apparently include a 
higher level of intrinsic variability than the narrow emission lines.  Spectropolarimetric 
observations spanning (and densely sampling) at least 2 complete rotation cycles are 
required to study in more details the relative strength of rotational modulation and 
intrinsic variability in most spectral lines of interest.  

Rotational modulation of photospheric and narrow emission lines can be mostly attributed 
to the presence of dark spots and hot accretion regions at the surface of BP~Tau. 
We find that both types of features grossly overlap, concentrate near the pole and cover 
altogether a relative area of about 2\%, similar to what was observed on V2129~Oph (D07).  
This agrees with previous independent observational and theoretical studies estimating 
that the accretion spots typically cover no more than a few percent of the stellar surface 
\citep[e.g.,][]{Valenti04, Jardine06}.  It also agrees with the findings of 
\citet{Valenti03} that BP~Tau (and other cTTSs) are faintest during times of maximum 
emission fluxes and polarisation signatures.  

Veiling is also detected in the photospheric spectra, at an absolute level varying between 
0 and 60\%, with relative changes of up to $\pm15$\% within a single run.  We find that 
veiling only weakly correlates with the amount of emission and polarisation, e.g., in \hei\ 
lines, and is subject to intrinsic fluctuations much larger than those seen in line 
emission fluxes and polarisation signatures.  This is similar to what \citet{Valenti03} 
report, strengthening their conclusion that veiling is rather due to stochastic variations 
in accretion rates rather than to the magnetic geometry and the location of accretion spots.  
It also suggests that veiling (as a pure continuum excess) is likely not causing 
the observed Stokes $I$ profile distortions in photospheric lines (as both would otherwise 
neatly correlate together), bringing further support 
to our model in which profile distortions of photospheric lines are mainly caused by 
dark surface spots.  

If our model is confirmed, it would indicate that that the heat produced in the accretion shock 
is not transferred to the photosphere efficiently enough to warm it up above the temperature 
of the non-accreting photosphere.  

\subsection{Magnetic topology and origin of the field}

Stokes $V$ signatures are detected on BP~Tau, both from photospheric and narrow emission 
lines;  the very different Zeeman signatures that both sets of lines exhibit (featuring in 
particular opposite line-of-sight polarities) suggest that they trace magnetic fields from 
different complementary regions of the stellar surface, with emission lines tracing the 
fields of accreting regions and photospheric lines tracing the fields of non-accreting 
regions (as in V2129~Oph, see D07).  With this model, we are able to recover the large-scale  
magnetic topology of BP~Tau from both sets of Zeeman signatures, and at both observing 
epochs.  The reconstructed magnetic field involves mainly a dominant dipole component 
of 1.2~kG and a strong octupolar term of 1.6~kG, both only slightly tilted with respect 
to the rotation axis.  

Zeeman signatures demonstrate that strong fields are present at the surface of BP~Tau.  
We find that the average magnetic filling factor, describing the relative amount of light 
from magnetic regions within each local surface area at the surface of BP~Tau, is about 
25\%;  in this context, the peak magnetic fluxes of 3~kG we reconstruct on BP~Tau indicate 
that magnetic fields likely reach maximum field strengths of up to 12~kG, mostly in accreting 
regions.  This result is supported by the strong longitudinal fields (in excess of 8~kG) 
traced with the \hei\ 667.815~nm line, presumably forming mostly in the magnetic accreting 
regions (with little contribution from adjacent non-magnetic areas) and thus less prone to 
longitudinal dilution than other narrow emission lines.  This is also in qualitative 
agreement with the study of magnetic broadening of infrared lines of BP~Tau \citet{Johns99b}, 
indicating that fields of up to 10~kG are likely to be present at the surface of the star.  
We note that the magnetic pressure within the accreting regions vastly exceeds the gas 
pressure of the surrounding non-magnetic photosphere, making it difficult to understand why 
and how accreting spots are spatially confined (even more so than in the case of
sunspots);  one possible (though fairly speculative) reason could be that magnetic fields 
are also present in the quiet photospheric regions of BP~Tau, but only in the form of 
small-scale tangled multipolar regions producing no detectable circular polarisation.    

Considering the whole surface of BP~Tau, we find that the average magnetic flux over 
non-accreting regions is about 1.2~kG while that over the complete star is 1.4~kG;  
looking now at the visible hemisphere only (e.g., at phase 0.0 in the Feb06 image), 
we find that the average magnetic flux from non-accreting regions is again about 1.2~kG 
whereas that over the full visible hemisphere is 1.8~kG.  The first of these 2 values is 
in good agreement with average magnetic fluxes derived from optical lines by 
\citet{Johns99b}.  These authors however find that infrared lines indicate larger 
average magnetic fluxes (of as much as 2.8~kG, though \citealt{Johns07} mentions 
2.2~kG only), presumably because these lines also 
trace magnetic fields in cool highly magnetic spots whose brightness contrast with 
respect to the photosphere is much smaller at infrared than at optical wavelengths.  
Although we also find that strongest magnetic fields indeed concentrate in cool spots, 
the average brightness-unweighted magnetic flux we derive (1.8~kG) is smaller 
than that estimated by \citet{Johns99b} and \citet{Johns07}, suggesting that we still 
likely miss magnetic flux from cool spots (e.g., from dark non-accreting small-scale 
magnetic regions).  
 
Comparing with V2129~Oph, we find that BP~Tau hosts a 4 times stronger dipole field 
and a half as strong toroidal field.  We speculate that this difference likely reflects  
the fact that BP~Tau is still fully-convective, while V2129~Oph had recently started 
to build up a radiative core (D07).  Fully convective stars are indeed much more successful  
at triggering strong nearly-axisymmetric low-order poloidal surface magnetic topologies 
\citep[][Morin et al., 2008 in preparation]{Donati06a, Hallinan06, Morin07, Hallinan07} than their 
more massive partly-convective counterparts 
\citep[e.g.,][]{Donati03}.  If confirmed, this result would suggest that the strong fields 
of low-mass cTTS are likely dynamo-generated rather than fossil fields;  the underlying processes 
capable of producing such fields almost without the help of differential rotation are not 
yet fully understood from a theoretical point of view \citep{Chabrier06, Dobler06}.  
In particular, it is not clear how dynamo processes are able to produce magnetic fields 
that vastly exceed the thermal equipartition value, such as those of BP~Tau and fully 
convective M dwarfs.  

The large-scale magnetic topology of BP~Tau has apparently undergone no more than 
small changes between Feb06 and Dec06 -- apart from a global phase shift of about 0.25 
rotation cycle that could be due to a slight error on the assumed rotation period.  
It suggests that the lifetime of BP~Tau's magnetic topology is 
comparable to the timespan between our two sets of observations, i.e., much longer than 
those of partly-convective active stars (whose field generally changes beyond recognition 
in no more than a few weeks, e.g., \citealt{Donati03}).  This result is in agreement 
with recent claims that magnetic fields of fully-convective main-sequence dwarfs are stable 
on timescales of at least 1~yr \citep[Morin et al., 2008 in preparation]{Morin07}.  
If confirmed, it would imply that the nominal rotation period of BP~Tau (7.6~d) is slightly 
underestimated;  a better phase match between both images (separated by 298~d or 39 rotation 
cycles) is obtained when assuming that the rotation period is 7.65~d.  Note that this new 
estimate is very close to the value of 7.64~d initially derived by \citet{Vrba86} from 
multicolour photometric data.  

\subsection{Disc-star magnetic coupling}

Several theoretical papers \citep[e.g.,][]{Konigl91, Shu94, Cameron93, Long05} studied 
how the stellar magnetic field interacts with the surrounding accretion disc
and disrupts its vertical structure close to the star.  They showed further that
the balance between accretion torques and angular momentum losses causes the
rotation of the star to evolve towards an equilibrium in which the disc
disruption radius lies close to \rmag\ and just inside the co-rotation radius 
\rcor.  They proposed that this coupling causes cTTSs to slow down to the  
Keplerian orbital period at a radius about 10--50\% larger than \rmag\ (i.e., 
$\rmag\simeq0.8$~\rcor), explaining why cTTSs are on average 
rotating more slowly than their disc-less equivalents.  This scenario is 
refered to as `disc-locking' in the literature.  

For BP~Tau, we obtain that the 1.2~kG large-scale dipole field we observe 
yields an equilibrium radius \rcor\ ranging between 6.3 and 8.9~\rstar\ (depending on whether 
magnetic diffusivity is due to buoyancy or turbulent diffusion) according to the model of 
\citet{Cameron93}.  
Similar results are obtained from the model of \citet{Long05}, yielding 
$\rcor=8.5$~\rstar\ and $\rmag=6$~\rstar\ when applied to BP~Tau.  
This is in agreement with both our own independent estimates of \rcor\ (about 7.5~\rstar) and 
\rmag\ (larger than 4~\rstar, see Sec.~\ref{sec:mag}).  
Using the model of \citet{Konigl91} however, 
we find an equilibrium radius of 2.7~\rstar, in strong contradiction with both 
our estimates of \rcor\ and \rmag.  As for V2129~Oph, we find that the disc-locking 
scenario is compatible with observations when the magnetic coupling between the star 
and its accretion disc is described by models like those of \citet{Cameron93} or 
\citet{Long05}.  

Our results suggest that winds on BP~Tau are obviously not strong enough to blow 
open field lines larger than 3~\rstar\ \citep[e.g.,][]{Safier98, Matt04} and to prevent 
disc magnetic disruption to take place and magnetospheric accretion to occur as far as 
\rmag.  If it were the case, accretion spots would distribute differently over the 
surface of the star, with a significant fraction of mass being accreted towards the 
equator, in strong contradiction with our observations.  Our observations also 
suggest that the accretion disc is magnetically warped at the vicinity of \rmag;  
a similar conclusion was reached in the case of the cTTS AA~Tau on completely 
different arguments \citep{Bouvier07b}.

\section{Conclusion}

In this paper, we report the detection of spectropolarimetric Zeeman signatures 
on the cTTS BP~Tau using mostly ESPaDOnS/CFHT (with help from NARVAL/TBL).  
Circular polarisation signatures in photospheric lines and in narrow emission 
lines tracing magnetospheric accretion are monitored throughout most of the 
rotation cycle of BP~Tau at two different epochs (Feb06 and Dec06).  We find 
that polarised and unpolarised spectral proxies tracing the photosphere 
and the footpoints of accretion funnels show temporal variations that are 
mostly attributable to rotational modulation.  
 
From our spectropolarimetric data sets of photospheric and narrow \caii\ emission 
lines simultaneously, we reconstruct, using tomographic imaging, the large-scale 
magnetic topology and the location of accretion spots at the surface of BP~Tau at 
both epochs.  We find that the magnetic topology of BP~Tau involves a dominant 
(1.2~kG) dipole but also a strong (1.6~kG) octupole, both slightly (though 
differently) tilted with respect to the rotation axis.  In particular, the 
strong dipole component makes BP~Tau fairly different from V2129~Oph, another 
(more massive) cTTS on which a similar study was carried out (D07).  
Accretion spots coincide with the two main high-latitude octupole poles and overlap 
with dark photospheric spots;  they each cover about 2\% of the stellar surface.

Despite clear variability between both epochs, the large-scale magnetic topologies 
we reconstruct are nevertheless grossly similar, suggesting an overall lifetime 
longer than 6 months.  The strong mostly-poloidal, nearly-axisymmetric field 
of BP~Tau (and in particular its long-lived dipole component) is very reminiscent of 
magnetic topologies of fully-convective dwarfs 
\citep[][Morin et al., 2008 in preparation]{Donati06a, Hallinan06, Morin07, Hallinan07}.  
It suggests that the strong large-scale poloidal fields hosted by fully-convective 
cTTSs such as BP~Tau (but absent in more massive non fully-convective cTTSs like 
V2129~Oph) are likely not fossil remants, but rather result from vigorous dynamo 
action operating within the bulk of their convective zones.  

Preliminary modelling suggests that the magnetosphere of BP~Tau 
must extend to distances of at least 4~\rstar\ to produce accretion spots at a 
latitude roughly matching those we observe.  At the very least, it demonstrates 
that magnetic field lines from the protostar are not blown open close to the surface 
by a stellar wind, but are apparently capable of coupling to the accretion disc beyond 
3~\rstar.  Our estimates of the magnetospheric and corotation radii for BP~Tau match 
the theoretical predictions of \citet{Cameron93} and \citet{Long05}, thus strengthening 
the idea that star/disc magnetic coupling may indeed be responsible for the slow 
rotation of fully-convective cTTSs such as BP~Tau.

\section*{Acknowledgements}

We thank the CFHT and TBL staff for their help during data collection.  We also 
thank the referee, C.~Johns-Krull, for valuable comments on the manuscript.

\bibliography{bptau}

\bibliographystyle{mn2e}

\end{document}